\documentclass[conference]{IEEEtran}
\IEEEoverridecommandlockouts
\usepackage{cite}
\usepackage{amsmath,amssymb,amsfonts}
\usepackage{algorithm}
\usepackage[noend]{algorithmic}
\usepackage{graphicx}
\usepackage{xcolor}
\usepackage{booktabs}
\usepackage{multirow}
\usepackage{hyperref}
\usepackage{caption}
\usepackage{array}
\usepackage{iftex}
\usepackage{subcaption}
\usepackage{enumitem}

\newcommand{\sysname}{{RcLLM}}

\def\BibTeX{{\rm B\kern-.05em{\sc i\kern-.025em b}\kern-.08em
    T\kern-.1667em\lower.7ex\hbox{E}\kern-.125emX}}

\begin{document}
\bstctlcite{BSTcontrol}


\title{RcLLM: Accelerating Generative Recommendation via Beyond-Prefix KV Caching}




\author{\IEEEauthorblockN{Zhan Zhao}
\IEEEauthorblockA{\textit{Department of Computer Science} \\
\textit{Hong Kong Baptist University}\\
Hong Kong, China \\
zhanzhao@comp.hkbu.edu.hk}
\and
\IEEEauthorblockN{Yuxin Wang}
\IEEEauthorblockA{\textit{Department of Computer Science} \\
\textit{Hong Kong Baptist University}\\
Hong Kong, China \\
yxwang@comp.hkbu.edu.hk}
\and
\IEEEauthorblockN{Amelie Chi Zhou}
\IEEEauthorblockA{\textit{Department of Computer Science} \\
\textit{Hong Kong Baptist University}\\
Hong Kong, China \\
amelieczhou@comp.hkbu.edu.hk}
}

\maketitle


\begin{abstract}


Large Language Models (LLMs) are transforming recommendation into a generative task, yet their industrial deployment is hindered by the prohibitive latency of processing long, personalized prompts where standard prefix caching fails due to non-contiguous reuse patterns.
To address this, we propose \sysname{}, a distributed inference system that accelerates generative recommendation via Beyond-Prefix KV Caching. \sysname{} decomposes prompts into reusable blocks and manages the massive scale of item catalogs through a stratified distributed storage architecture: \emph{compact user histories} are replicated for zero-latency retrieval, while \emph{massive item caches} are sharded using a similarity-aware placement algorithm. 
\sysname{} effectively eliminates redundant quadratic computation with two strategic techniques: 1) an \emph{affinity-based global scheduler} to maximize data locality, and 2) a \emph{selective attention} mechanism to correct approximation errors.
Evaluations on real-world datasets demonstrate that \sysname{} reduces Time-To-First-Token (TTFT) by 1.31$\times$–9.51$\times$ compared to state-of-the-art prefix caching systems, enabling real-time serving compliance with negligible impact on recommendation accuracy.

\end{abstract}


\section{Introduction}


Large Language Models (LLMs) are transforming recommendation systems from static retrieval engines~\cite{updlrm2024,yu2026nearzero} into generative reasoners. Unlike classical ID-based pipelines, LLM-based recommenders~\cite{zhou2025onerectechnicalreport,li2023gpt4recgenerativeframeworkpersonalized} can rank or generate items by reading rich user contexts (e.g., interaction traces and natural-language preferences) and reasoning deeply over item semantics. However, while this unification of heterogeneous signals enhances explainability and personalization, it introduces severe serving costs that clash with the strict latency Service Level Objectives (SLOs) of high-traffic industrial environments~\cite{yu2026nearzero}.

For instance, industrial stacks like Amazon's often require end-to-end response latencies within $\sim$100 ms~\cite{aws2021recsys1,gupta2020deeprecsys}, while the model-side budget is typically only a small fraction of that after accounting for network hops, feature retrieval, ranking orchestration, and business logic. This constraint is especially severe for long-context LLM inference. In generative recommendation, prompts typically aggregate extensive user histories and verbose item metadata (e.g., titles and descriptions), often spanning thousands of tokens~\cite{li2023gpt4recgenerativeframeworkpersonalized}.


The computational cost of this extensive context is dominated by the prefill stage~\cite{jiang2024minference}, which constructs the Key-Value (KV) cache for the prompt~\cite{wolf2020transformers}. Due to the quadratic complexity of self-attention, prefill latency scales poorly with prompt length. Crucially, recommendation-style generation is structurally distinct from conversational tasks: it is ``prefill-heavy and decode-light''~\cite{li2023gpt4recgenerativeframeworkpersonalized}, typically emitting only a ranked list of items or short IDs. Consequently, the massive prefill overhead cannot be amortized over a long decoding phase, making Time-To-First-Token (TTFT) the definitive bottleneck in the serving pipeline.

To mitigate prefill costs, modern LLM serving stacks rely heavily on prefix caching~\cite{vllm2023,zheng2024sglang}, which reuses KV states for requests sharing identical leading segments. While effective for chatbots, this paradigm is structurally incompatible with recommendation workloads where the most computationally expensive segments rarely form a stable prefix. Specifically: (i) \textbf{user histories}, though semantically redundant across the population, are lexically unique and positionally fluid for each user; and (ii) \textbf{candidate item sets} are frequently permuted and appended after these variable user contexts.
Because conventional prefix caches require a contiguous match from the first token, they suffer from near-zero hit rates on this dominant token mass, forcing redundant $O(n^{2})$ attention for the same content and leading to tail-latency violations that fail to meet stringent industrial recommendation SLOs~\cite{aws2021recsys1,gupta2020deeprecsys}.

To address this challenge, we decompose LLM recommendation prompts into two reusable KV components that enable \textbf{beyond-prefix reuse}: (i) a semantic history KV cache pool that captures and quantizes recurring patterns in user reviews and interactions, and (ii) a candidate item KV cache pool that precomputes and shards item-level KV blocks offline for non-contiguous assembly during online prefill.

While this decomposition enables beyond-prefix reuse, implementing it at scale presents significant system challenges: \emph{First}, the item catalog is massive, requiring a distributed KV storage strategy that balances limited per-node capacity with high cache hit rates. \emph{Second}, distributing KV states across nodes introduces non-trivial scheduling challenges, as poorly coordinated execution can place cross-node communication on the critical path of inference. \emph{Finally}, stitching together non-contiguous KV blocks disrupts the model's global attention, potentially degrading recommendation accuracy.

To surmount these hurdles, we present \textbf{\sysname{}}, a distributed inference system that decouples KV management from online execution via a split-phase architecture. Specifically, at \textbf{offline phase}, \sysname{} (1) builds a compact, replicated semantic history library to handle recurring user review patterns, and (2) shards the massive item KV cache across the cluster using a similarity-aware placement strategy that preserves locality. At the \textbf{online phase}, a cache-aware scheduler routes requests to nodes with high cache affinity while balancing queueing loads. The local execution engine then performs zero-copy assembly of non-contiguous KV blocks. To mitigate approximation errors from block reuse, \sysname{} employs selective recomputation: it identifies and recomputes a small set of ``heavy-hitter'' tokens that disproportionately influence attention, while reusing cached states for the remaining majority.

We evaluate \sysname{} on Amazon, Yelp, and Goodreads datasets. \sysname{} substantially improves TTFT compared to prefix caching, delivering a $1.31\times$--$9.51\times$ speedup while maintaining competitive ranking quality under appropriate recomputation budgets. The reduction in TTFT makes it feasible for LLM-based recommendation systems to operate within stringent industrial tail-latency SLOs, bridging the gap between model capability and production requirements.

In summary, we make the following contributions:
\begin{itemize}
    \item We identify and quantify significant beyond-prefix redundancy in LLM-based recommendation prompts, concentrated in (i) static candidate item descriptions and (ii) semantically repetitive user histories, making prefill computation highly redundant.
    \item We design \sysname{}, a distributed serving architecture that enables cross-request KV reuse for non-contiguous, beyond-prefix segments via a stratified cache (semantic-history pool + sharded item-KV pool), together with similarity/graph-aware cache placement and cache-aware scheduling.
    \item We demonstrate that RcLLM delivers up to 9.51$\times$ speedups over prefix caching while preserving ranking accuracy, bridging the gap between LLM capability and industrial latency requirements.
\end{itemize}
\section{Background and Motivation}


\subsection{LLM-Based Recommendation Systems}\label{sec:2a}





The landscape of Recommendation Systems (RecSys) is undergoing a fundamental paradigm shift. While traditional approaches like the Deep Learning Recommendation Model (DLRM)~\cite{yu2026nearzero} rely on sparse ID-based features to model user preferences, recent advancements have moved toward prompt-based Large Language Models (LLMs). Emerging frameworks such as OneRec~\cite{zhou2025onerectechnicalreport} and HLLM~\cite{chen2024hllmenhancingsequentialrecommendations} demonstrate that leveraging the extensive world knowledge and reasoning capabilities of pre-trained models can significantly enhance recommendation accuracy. In these scenarios, the recommendation task is reformulated from a classification problem into a language generation or ranking task, where the model must ``read'' a user's context and ``reason'' about the best next item.


\begin{figure*}[htbp]
  \centering
  \begin{subfigure}[b]{0.29\textwidth}
    \centering
    \includegraphics[width=1.05\linewidth]{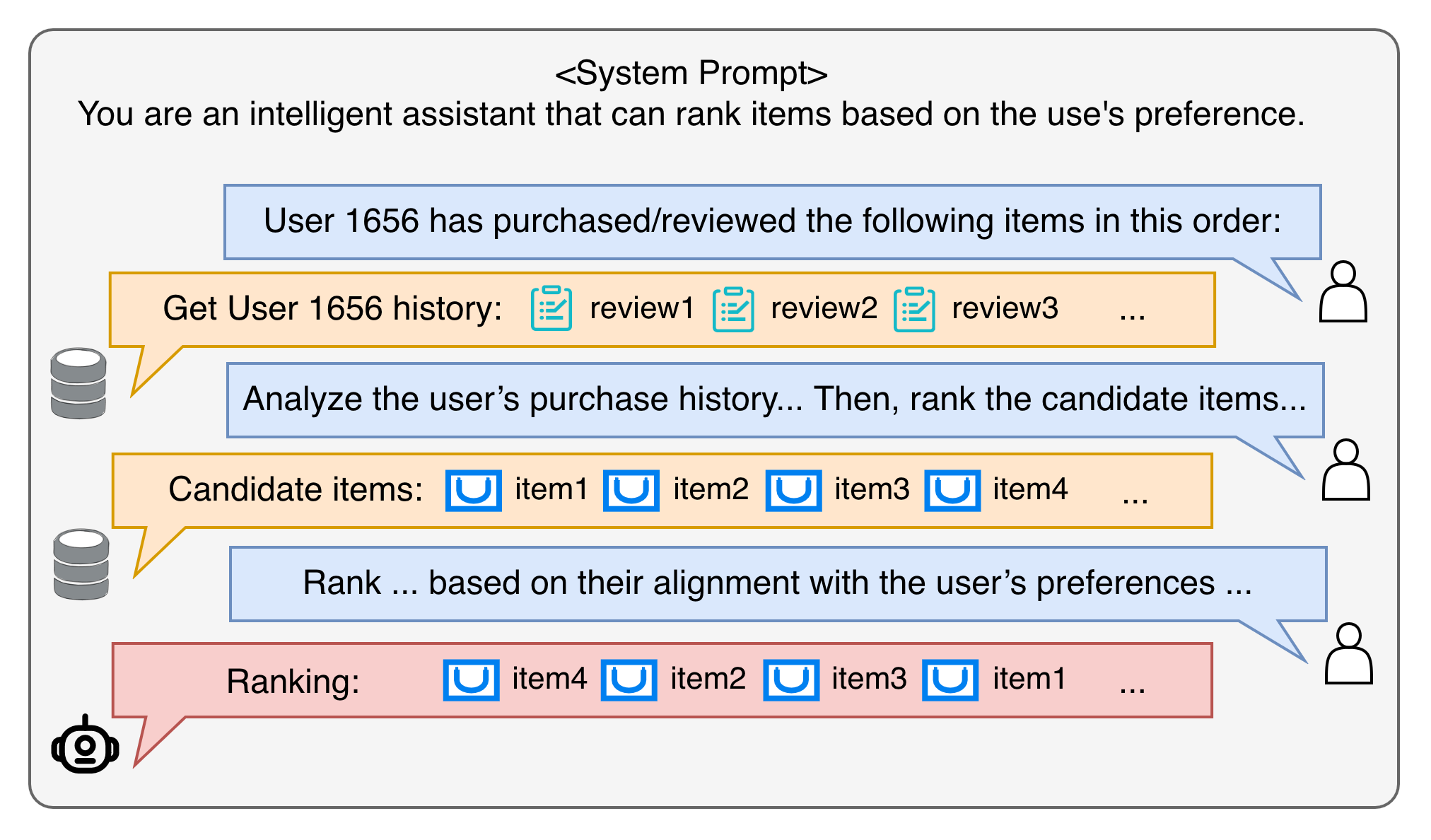}
    \caption{\scriptsize\bf Prompt Example}
    \label{fig:prompt-example}
  \end{subfigure}
  \hfill
  \begin{subfigure}[b]{0.38\textwidth}
    \centering
    \includegraphics[width=1.05\linewidth]{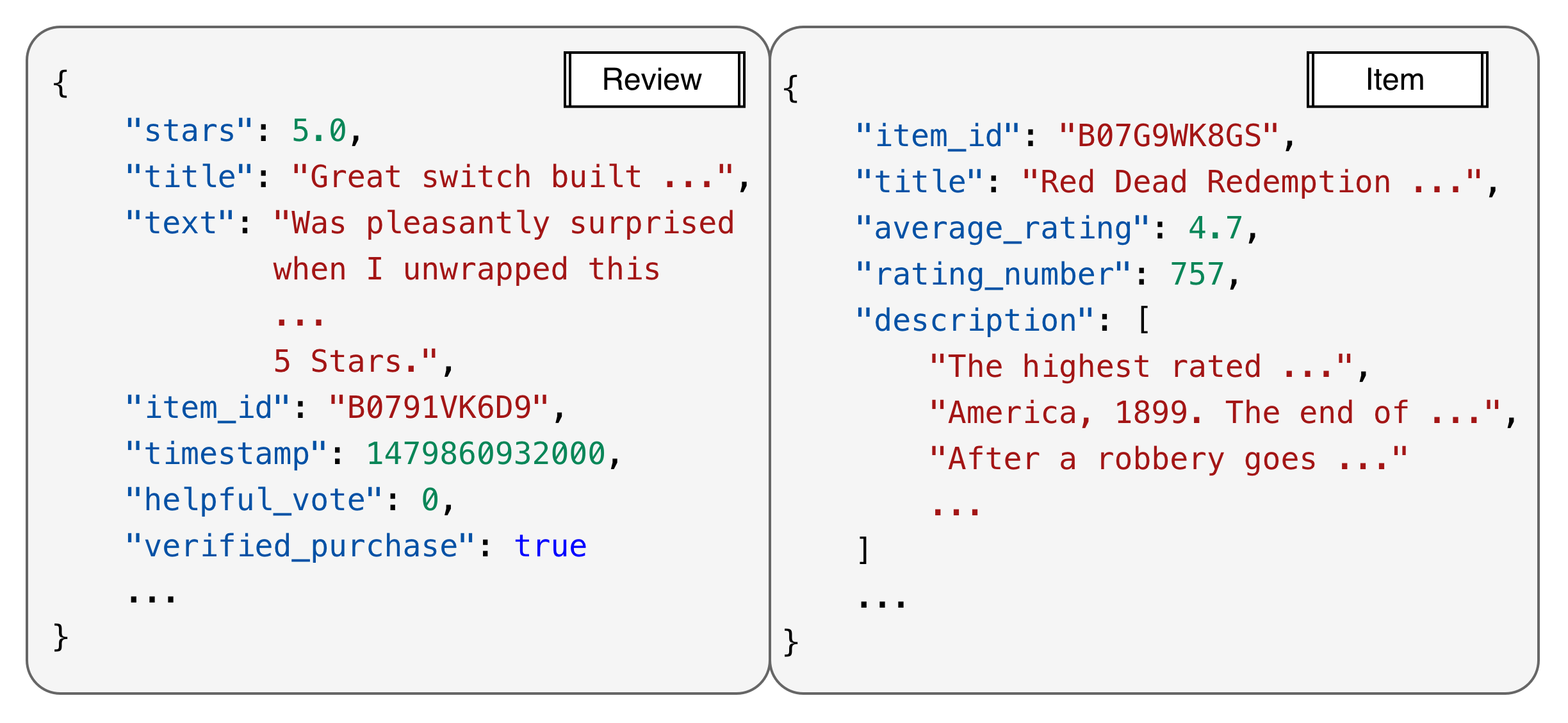}
    \caption{\scriptsize\bf Review and Item Example}
    \label{fig:item-example}
  \end{subfigure}
  \hfill
  \begin{subfigure}[b]{0.26\textwidth}
    \centering
    \includegraphics[width=1.05\linewidth]{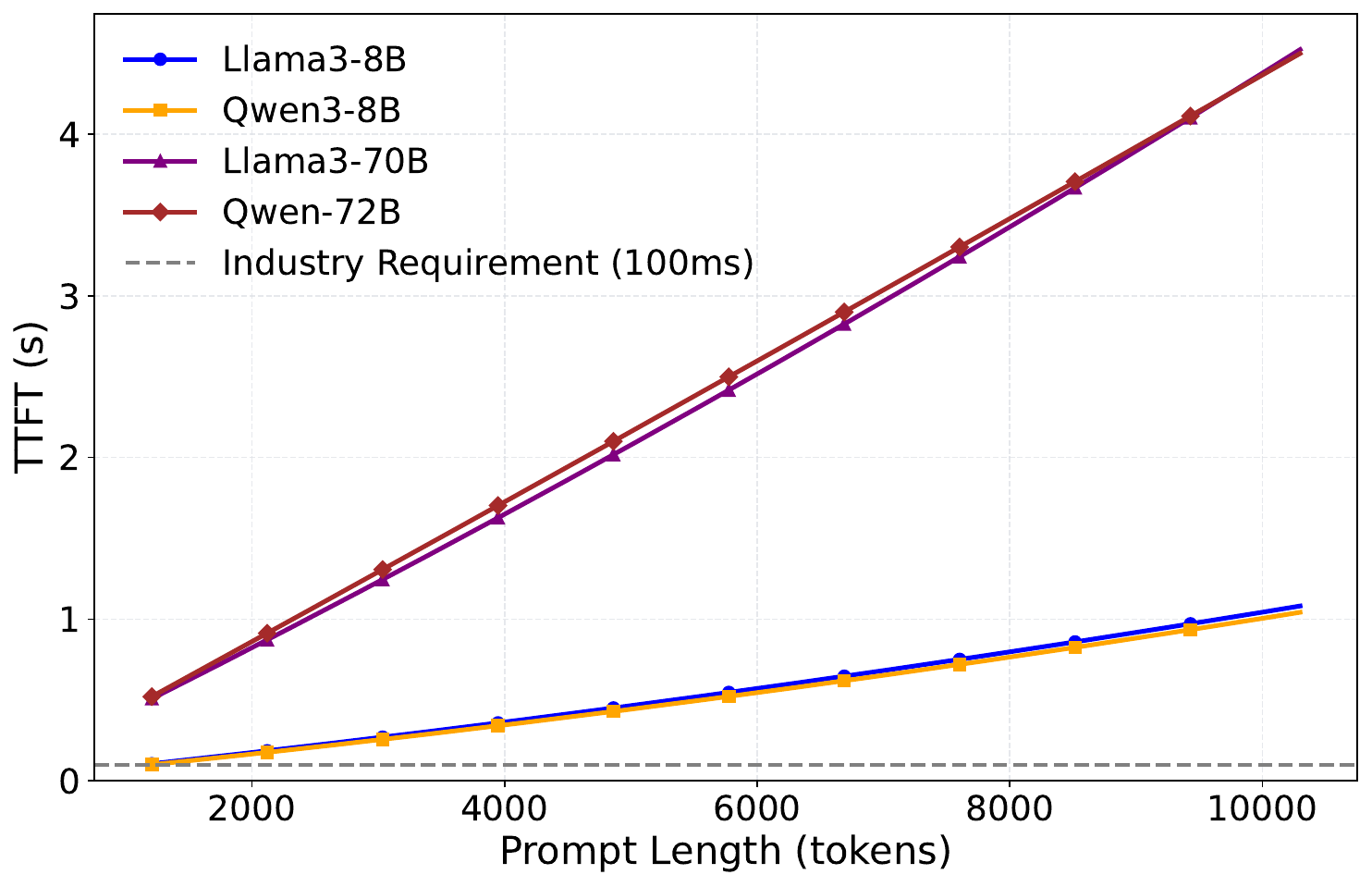}
    \caption{\scriptsize\bf Latency vs. Industrial SLO}
    \label{fig:latency-test}
  \end{subfigure}
  
  \caption{Prompt analysis for generative recommendation.
  (a) An example prompt combining user interaction history, candidate items, and task instructions.
  (b) An example item with typical product fields (title, category, description) forming mostly static prompt content, and a review example with semantically redundant text.
  (c) Inference latency exceeding industrial SLOs (typically under 100\,ms) as prompt length increases, underscoring the need for optimization.}
  \label{fig:prompt-analysis}
\end{figure*}

The core of this workflow is the prompt construction, which serves as the input workload for the model. As illustrated in Figure~\ref{fig:prompt-analysis}, the system does not ingest simple vectors, but rather constructs a composite textual prompt comprising three primary elements: detailed user history, a set of candidate items, and specific task instructions. 
For example, Figure~\ref{fig:prompt-example} illustrates the prompt structure used by an LLM-driven agent. The prompt begins with a system prompt specifying the agent’s objectives, followed by user-related information such as historical purchase or review records to capture personalized preferences. 
Afterward, a collection of candidate items is appended, where each item includes verbose attributes (titles, descriptions, metadata as shown in Figure~\ref{fig:item-example}) alongside detailed instructions guiding the ranking process. Leveraging this extensive contextual information, the LLM infers the user’s preferences and produces an ordered ranking of the items, which constitutes the final recommendation result.

To achieve high accuracy, these prompts act as extensive context windows, often exceeding thousands of tokens to capture the necessary semantic nuance. While this length is manageable for offline processing, it is prohibitive for real-time inference. In production recommender stacks, the model-side budget is typically very tight (e.g., $<100$\,ms~\cite{aws2021recsys1, gupta2020deeprecsys}), making the \emph{Time-To-First-Token (TTFT)} (i.e., time to emit the first recommendation item) the dominant latency metric. As shown in Figure~\ref{fig:latency-test}, TTFT increases sharply with prompt length on the Amazon Reviews dataset~\cite{mcauley2016amazon}, and can easily exceed this budget even with an 8B model. \textbf{This gap motivates system techniques that reduce the prefill cost on long, personalized recommendation prompts.} We next examine the underlying autoregressive inference mechanism.


\subsection{The Autoregressive LLM Inference}\label{sec:2b}

\begin{figure}[t]
  \centering
  \includegraphics[width=0.9\linewidth]{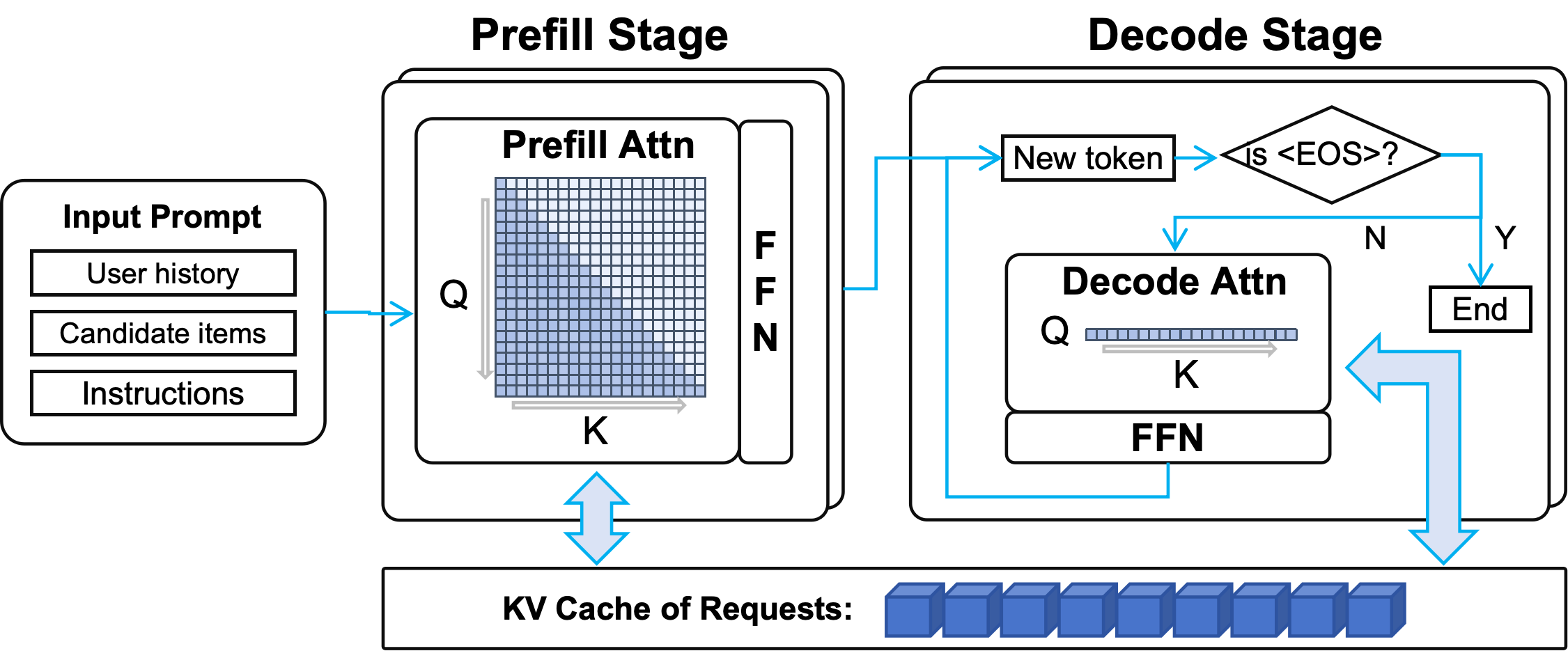}%
  \caption{Standard Autoregressive Inference Process}
  \label{fig:transformer-analysis}
\end{figure}


Modern LLM-based recommendation systems typically rely on Transformer-based architectures (e.g., Llama~\cite{grattafiori2024llama3herdmodels}, Qwen~\cite{yang2025qwen3technicalreport}) to generate rankings autoregressively. As illustrated in Figure~\ref{fig:transformer-analysis}, the inference workflow is functionally divided into two distinct phases, each with unique computational characteristics: the \emph{Prefill} stage and the \emph{Decode} stage: 
In the prefill phase, the model processes the entire prompt (user history, candidate items, and system instructions) in parallel to compute attention scores and populate the Key-Value (KV) cache for every token in the input sequence.
The core computational engine driving this stage is the \emph{Self-Attention} mechanism. For a sequence of length $n$, the mechanism projects tokens into Query ($Q$), Key ($K$), and Value ($V$) matrices and computes interactions via:
\begin{equation}\label{eq:attention}
  \text{Attention}(Q, K, V) = \text{softmax}\left(\frac{QK^T}{\sqrt{d_k}}\right)V
\end{equation}
where $d_k$ is the dimension of the key vectors. 
Once the prefill is complete and the first output token is generated, the system shifts to the decode stage. Here, tokens are generated sequentially. Each newly generated token is appended to the sequence and attended to the cached states until an end-of-sequence token is reached.

This architecture reveals the root cause of the latency bottleneck identified in Section~\ref{sec:2a}. The time complexity of the self-attention mechanism is $O(n^{2})$ with respect to the sequence length $n$. 
Since recommendation prompts are dominated by long interaction histories and verbose item descriptions (often exceeding thousands of tokens), the prefill stage bears the vast majority of the computational load and latency.
While state-of-the-art inference engines employ techniques like Prefix Caching~\cite{vllm2023,zheng2024sglang} to reduce prefill costs, they are structurally ill-suited for this workload. Traditional prefix caching is designed to reuse the initial static segment of a prompt (e.g., system instructions) within a single session or for requests sharing a common start. However, recommendation workloads require cross-request reuse for content that is not a prefix: identical candidate items appear repeatedly across millions of different requests, but they are often positioned after unique user histories or permuted in different orders. Because these repeated item blocks do not form a consistent prefix, standard mechanisms fail to detect them, forcing the engine to re-compute the massive KV cache from scratch for every requests. This structural inefficiency creates the primary barrier to real-time deployment of LLM-based recommendation systems.

\subsection{Insights and Motivation}\label{sec:motivation}

The severe latency bottleneck imposed by prefill stems from the inefficiency of treating every recommendation prompt as a unique computational event. In reality, the prompt context is dominated by two highly repetitive components:
1) \emph{candidate items}, which are immutable strings known offline, and 2) \emph{user histories}, which are sequences of past product reviews sharing a limited semantic vocabulary. Because these massive segments do not fundamentally change between requests, the corresponding $O(n^2)$ matrix calculations performed during the prefill stage are effectively redundant.
To validate the feasibility of eliminating this redundancy, we analyzed the data characteristics and attention mechanisms of representative LLMs (Qwen3-8B and Llama-3.1) on the Amazon Reviews dataset~\cite{mcauley2016amazon}. Our investigation reveals three critical insights:


\begin{figure*}[t]
  \centering
  \begin{subfigure}[b]{0.25\textwidth}
    \centering
    \includegraphics[width=1.05\textwidth]{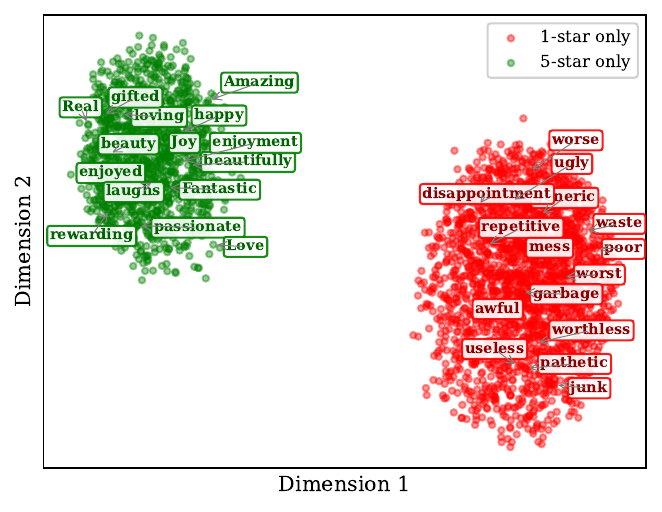}
    \vspace{-3ex}
    \caption{\scriptsize \textbf{Semantic Clustering of Tokens}}
    \label{fig:semantic-cluster}
  \end{subfigure}
  \hfil
  \begin{subfigure}[b]{0.25\textwidth}
    \centering
    \includegraphics[width=1.1\textwidth]{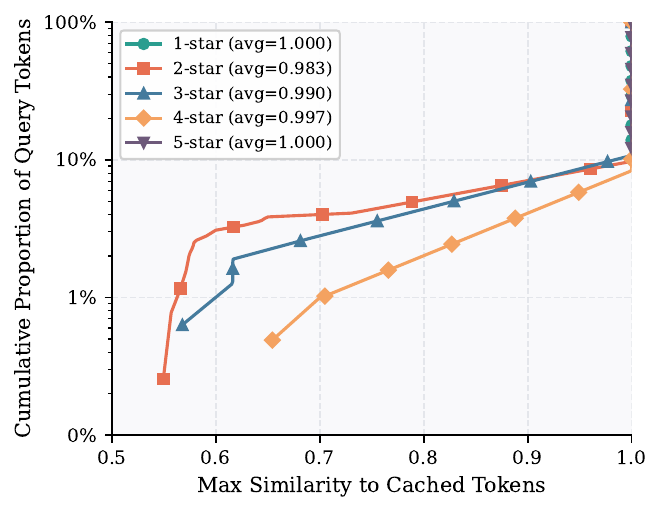}    \vspace{-3ex}

    \caption{\scriptsize \textbf{Token Reusability CDF}}
    \label{fig:embedding-cdf}
  \end{subfigure}
  \hfil
  \begin{subfigure}[b]{0.25\textwidth}
    \centering
    \includegraphics[width=1.05\textwidth]{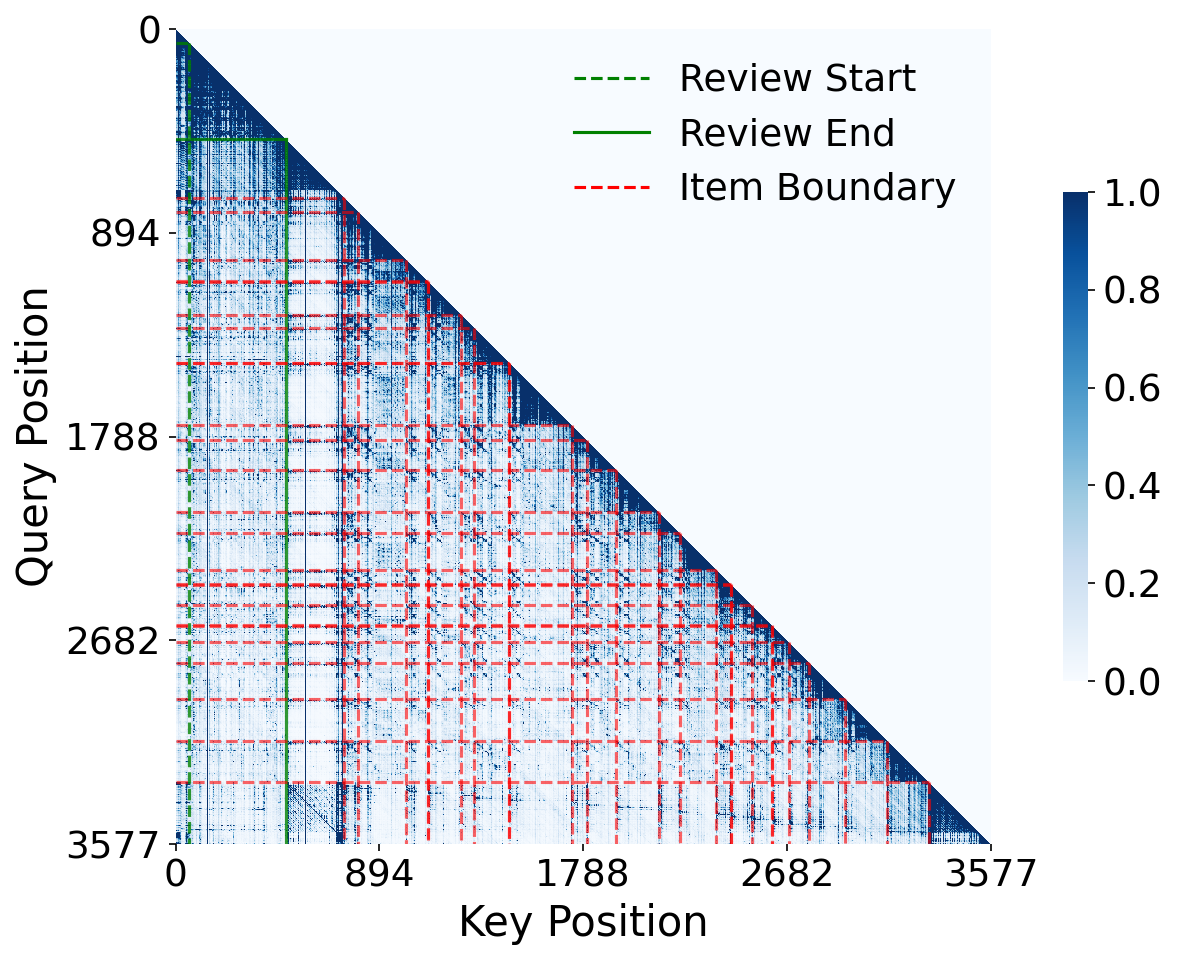}    \vspace{-3ex}
    \caption{\scriptsize \textbf{Attention Sparsity in Qwen3-8B}}
    \label{fig:attention-score}
  \end{subfigure}
  \vspace{-1ex}
  \caption{Analysis of token characteristics and attention mechanisms.
  (a) Visualizing token embeddings from 1-star and 5-star reviews in 2D space reveals distinct semantic clusters indicating a limited reusable vocabulary.
  (b) The CDF of cosine similarity shows over 93\% of tokens in new reviews find a near-identical match in a static historical pool, suggesting high potential for KV cache reuse.
  (c) The average attention score heatmap across heads and layers exhibits strong diagonal locality and sparsity, indicating minimal contribution from distant tokens absent specific \textbf{heavy hitters}.}
  \label{fig:analysis-overview}
\end{figure*}

\textbf{Insight 1: Semantic Redundancy in User History.}
User histories also occupy a significant portion of the prompt context and exhibit strong semantic locality. Unlike arbitrary dialogue, these reviews exhibit strong semantic patterns.
As shown in Figure~\ref{fig:semantic-cluster}, when we project token embeddings from 1-star and 5-star reviews, distinct clusters emerge, confirming that sentiments and descriptive patterns are highly consistent within specific rating groups. 
To quantify this reusability, we constructed a compact semantic library (on the order of $10^5$ position-aware prototypes) derived from historical data, and for each incoming history token we retrieve its nearest prototype in embedding space. Figure~\ref{fig:embedding-cdf} presents the Cumulative Distribution Function (CDF) of the resulting maximum cosine similarity scores. The results are compelling: more than 93\% of tokens in new user reviews find a near-identical match (cosine similarity close to 1.0) within this static pool. This demonstrates that the expensive prefill computation for user history is largely repetitive and can be effectively accelerated via semantic reuse.

\textbf{Insight 2: Static Nature of Candidate Items.}
Recommendation prompts include a substantial segment dedicated to candidate items, containing verbose attributes like titles and descriptions (see examples in Figure~\ref{fig:item-example}). Crucially, this information is \emph{inherently static} and fully known prior to inference. Consequently, the KV cache for millions of items can be pre-computed offline. This implies that instead of processing item descriptions repeatedly for every request, the system can simply retrieve and concatenate these static blocks during online inference.

\textbf{Insight 3: Sparsity and Locality in Attention.}
To ensure that aggressive caching and reuse strategies do not degrade ranking accuracy, we further analyze the self-attention patterns of Qwen3-8B. Figure~\ref{fig:attention-score} visualizes the average attention weights aggregated across all layers and heads. We observe two prominent characteristics.


\begin{itemize}[leftmargin=*]
    \item \textbf{Locality (Block-Diagonal Structure).} 
The attention heatmap exhibits strong diagonal-dominated structures, characterized by distinct triangular blocks aligned along the main diagonal. These blocks correspond exactly to the boundaries of the static candidate items described in Insight 1, confirming that the model processes each item as a self-contained semantic unit: tokens within an item description heavily attend to their neighbors (the triangle) but exhibit negligible cross-attention to the content of other items. 
\emph{This structural independence scientifically permits us to cache items as independent blocks and stitch them together without disrupting the model's reasoning.}
\item \textbf{Sparsity with Structural Anchors.}
Beyond local neighborhoods, long-range attention is highly sparse, with most distant token pairs receiving near-zero weights. However, the heatmap also reveals prominent vertical stripes corresponding to a small set of ``\emph{heavy hitter}'' tokens (e.g., system instructions or item delimiters) that consistently attract attention across the entire sequence. These tokens serve as global anchors, aggregating high-level semantic signals. 
\emph{This suggests that precise recomputation is only necessary for these few structural anchors, while the repetitive semantic content of the history can be safely retrieved from the semantic library.}
\end{itemize}

\textbf{Motivation.}
Collectively, these insights reveal a fundamental inefficiency in current serving paradigms: the prohibitive latency of LLM-based recommendation is not an intrinsic property of the task, but a consequence of failing to exploit its structural decomposability.
The \emph{block-diagonal attention} observed in Insight 3 validates that static candidate items (Insight 1) function as independent semantic units, permitting them to be cached as discrete blocks. Simultaneously, the presence of \emph{global anchors} (Insight 3) suggests that the semantic redundancy of user history (Insight 2) can be managed by preserving just a few critical tokens, rather than recomputing the entire sequence.
%
However, translating these insights into a practical system faces three fundamental challenges that standard inference engines cannot address:
\begin{itemize}[leftmargin=*]
    \item \textbf{Incompatibility with prefix caching}: Standard engines rely on requests sharing a common prefix. In recommendation, however, static candidate items are permuted and positioned after unique user histories, preventing standard reuse.
    \item \textbf{Massive storage}: Caching an industrial-scale item catalog requires terabytes of memory (e.g., Table~\ref{tab:kv-cache-scales-analysis}), far exceeding single-node GPU capacity.
    \item \textbf{Accuracy degradation}: Naively stitching together independent cached blocks disrupts the attention mechanism’s positional encoding, leading to severe ranking errors.
\end{itemize}


Driven by these limitations, we propose \emph{\sysname{}}, a specialized inference engine designed to bridge the gap.

\section{System Design}
\label{sec:system-design}

\subsection{System Overview}
\label{sec:design-overview}

\begin{figure}[t]
  \centering
  \vspace{-1ex}
  \includegraphics[width=0.9\linewidth]{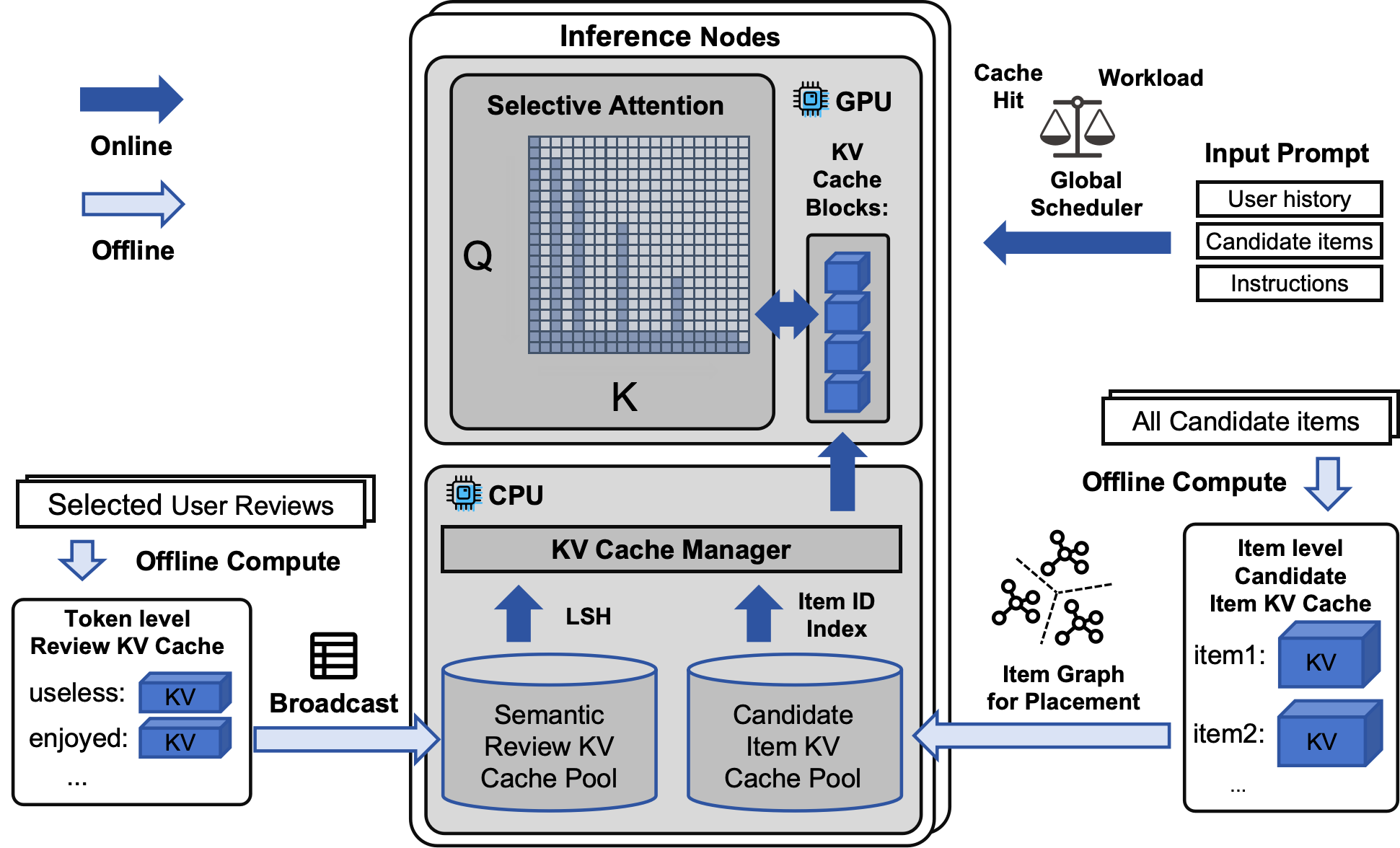}
  \caption{\sysname{} System Architecture
  }
  \label{fig:system-overview}
\end{figure}

In this paper, we propose \sysname{}, a distributed inference system designed to bridge the gap between the computational demands of generative recommendation and the strict latency constraints of industrial serving. As illustrated in Figure~\ref{fig:system-overview}, \sysname{} adopts a split-phase architecture that fundamentally decouples the management of KV cache states from real-time inference execution.

\subsubsection{Offline Phase (\S\ref{sec:offline-phase})}
The offline pipeline (bottom of Figure~\ref{fig:system-overview}) is designed to align storage strategies with the distinct scale and access patterns of recommendation data. Instead of a monolithic cache, \sysname{} stratifies the prompt context into two distinct KV storage pools as shown in Table~\ref{tab:kv-cache-summary}:
\begin{itemize}[leftmargin=*]
    \item \emph{Semantic History (Replicated)}: Recognizing that user histories are semantically repetitive but compact, we compress them into a \emph{semantic review library} and replicate it across all nodes. This ensures that the highly variable history component is always ``local'', eliminating network latency for this segment.
    \item \emph{Candidate Items (Partitioned)}: For the massive, immutable item catalog, we employ \emph{graph-based partitioning} to minimize cross-node KV retrieval overhead. By analyzing co-occurrence patterns, we identify clusters of correlated items and shard them onto specific nodes. This transforms the distributed storage problem into a locality optimization problem, proactively maximizing data locality for requests.
\end{itemize}

\begin{table}[t]
  \centering
  \scriptsize
  \caption{Summary of \sysname{}'s two cache pools}
  \label{tab:kv-cache-summary}
  \begin{tabular}{l c c c}
    \toprule
    \textbf{Cache Type} & \textbf{Scale} & \textbf{Accuracy} & \textbf{Placement} \\
    \midrule
    Semantic History & Small (GBs) & Approx. & Replicated \\
    \addlinespace
    Candidate Item & Massive (TBs) & Exact & Sharded \\
    \bottomrule
  \end{tabular}
\end{table}

\subsubsection{Online Phase (\S\ref{sec:online-phase})} 

In standard LLM serving, the scheduler treats nodes as identical and routes requests using simple stateless algorithms (e.g., round robin). Some prefix-aware scheduling algorithms\cite{srivatsa2025preble} may also degrade due to too few prefix match opportunities.
However, in \sysname{}, the massive item cache is sharded across nodes and worker nodes become ``stateful'' (holding different item shards).
The online phase shifts the scheduling paradigm from stateless load balancing to stateful, locality-aware orchestration:
\begin{itemize}[leftmargin=*]
\item \emph{Locality-Aware Routing}: The \emph{global scheduler} routes requests based on cache affinity, namely directing computation to the specific node where the required item KV blocks are resident, effectively ``moving compute to data''.

\item \emph{Zero-Copy Execution}: Once at the target node, the engine addresses non-prefix access patterns via \emph{zero-copy block retrieval}. Instead of physical concatenation, it maps logical prompt sequences to scattered physical memory pages. Finally, it employs \emph{selective attention} to dynamically stitch these blocks together, correcting boundary artifacts to preserve model fidelity.
\end{itemize}



In the following, we introduce technical details of the two phases separately.

\subsection{Offline Phase: Distributed KV Construction and Placement}
\label{sec:offline-phase}

The offline phase is responsible for materializing reusable KV states and preparing them for efficient access during online inference. As introduced above, \sysname{} builds two fundamentally different KV cache pools, including  the KV cache \emph{pool for semantic review histories} and the \emph{pool for candidate items}.

\textbf{First, for semantic review histories}, the cache must satisfy three key requirements: (i) \emph{Compact}, as user histories are included in every request and accessed frequently; (ii) \emph{Position-aware}, since Transformer representations couple token semantics with positional information; (iii) \emph{Low-latency} local access, to avoid cross‑node communication on the critical inference path.
%
%
We construct the semantic history KV cache using a three-stage pipeline to meet these design requirements: 

\begin{itemize}[leftmargin=*]
    \item \emph{Position-Aware Embedding}: To preserve positional effects, \sysname{} embeds each token jointly with its position. For a token $t$ appearing at position $p$, its embedding $e_{t,p}$ incorporates both lexical semantics and positional encoding. This avoids incorrect reuse across tokens that are semantically similar but appear in different structural contexts.

\item \emph{LSH-Based Semantic Clustering}: 
The position-aware embeddings are clustered using Locality-Sensitive Hashing (LSH)~\cite{jafari2021survey}, which efficiently groups tokens into discrete semantic regions. This step quantizes the large space of user history tokens into a bounded set of semantic prototypes, each representing a recurring semantic–positional pattern.

\item \emph{KV Materialization}: 
For each semantic prototype, \sysname{} selects a canonical representative token (e.g., the cluster centroid) and pre-computes its layer-wise KV states offline. These precomputed KV blocks form the semantic history cache and can be reused across requests via nearest-prototype matching at inference time. In our implementation, this process yields a compact cache containing on the order of $10^5$ semantic prototypes.
Though compact, the size of this cache remains too large to reside entirely in GPU memory (e.g., $\sim$30GB for $10^5$ prototypes using Qwen3-8B model).
\sysname{} therefore stores it in CPU memory and replicates it across all serving nodes. 
\end{itemize}

Our semantic clustering introduces approximation into the KV states, and could hence harm the accuracy of inference.
\sysname{} mitigates approximation errors via selective recomputation during online inference (\S\ref{sec:online-phase}), which recomputes a small set of structurally important tokens (heavy hitters) while reusing the remaining majority.

\begin{figure}[t]
  \centering
  \begin{minipage}[t]{0.48\linewidth}
    \vspace{0pt}
    \centering
    \scriptsize
    \setlength{\extrarowheight}{2pt} 
    \resizebox{\linewidth}{!}{%
        \begin{tabular}{| p{2em} | p{4em} | p{4em} | p{4em} |}
          \hline 
          
          \multirow{2}{=}{\textbf{Item Count}} & \multicolumn{3}{c|}{\textbf{Token Count per Item}} \\ 
          
          \cline{2-4} 
          
          & 50 & 100 & 200 \\
          \hline 
          
          10K   & 137.5\,GB & 275\,GB   & 550\,GB \\
          100K  & 1.38\,TB  & 2.75\,TB  & 5.5\,TB \\
          200K  & 2.75\,TB  & 5.5\,TB   & 11\,TB \\
          500K  & 6.88\,TB  & 13.75\,TB & 27.5\,TB \\
          1000K & 13.75\,TB & 27.5\,TB  & 55\,TB \\
          \hline 
        \end{tabular}%
    }
    \vspace{1ex}
    \captionof{table}{Item KV Cache Scale Analysis (Qwen3-8B)}
    \label{tab:kv-cache-scales-analysis}
  \end{minipage}%
  \hfill
  \begin{minipage}[t]{0.48\linewidth}
    \vspace{-2pt}
    \centering
    \includegraphics[width=1.05\linewidth]{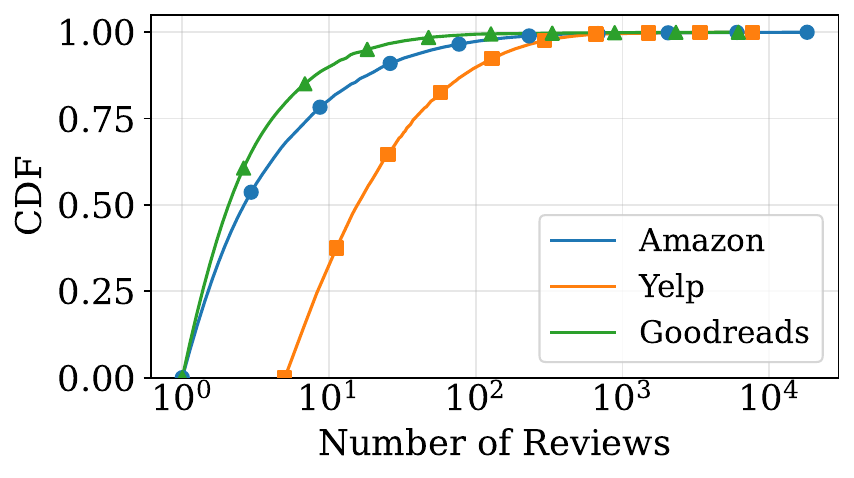}
    \vspace{-4ex}
    \captionof{figure}{Item popularity distribution of three datasets}
    \label{fig:item-popular}
  \end{minipage}
\end{figure}

\textbf{Second, for candidate items}, 
\sysname{} precomputes their KV blocks offline and reuses them as non-contiguous blocks during online prefill, adjusting positions at assembly time. Any cross-context discrepancy introduced by block reuse is mitigated by selective recomputation (\S\ref{sec:online-phase}).

The candidate item catalog presents a scale challenge. To understand the scale of these KV states, we analyzed three datasets: Amazon Reviews \cite{hou2024bridginglanguageitemsretrieval}, Yelp \cite{yelp2025} and Goodreads \cite{wan2018item}.
The total number of items is high, e.g., reaching 778K in Amazon for just three categories. The average length of their candidate items are 87, 76 and 124, respectively. The storage requirement for caching all item KVs in the three datasets is shown in Table~\ref{tab:kv-cache-scales-analysis}, which easily reaches terabytes in size. 
Since item
descriptions are immutable, we pre-compute the KV matrices for all candidate items and treat the precomputed data as a sharded distributed KV store.
Unlike the replicated semantic history library, a specific item's KV block exists only on a subset of nodes, necessitating a placement strategy that minimizes remote fetches.

The primary objective of our \emph{placement strategy} is to maximize the local cache hit rate. Achieving this requires navigating two conflicting data characteristics, namely \emph{popularity skew} and \emph{item correlation}. As illustrated in Figure~\ref{fig:item-popular}, item popularity follows a heavy-tailed distribution. Naively partitioning hot items to a single node would create ``hotspots" and frequent cache misses for other nodes. 
On another hand, items often exhibit strong co-occurrence patterns (e.g., books in a specific series). Placing highly relevant items on the same instance minimizes the network overhead of fetching scattered KV blocks, treating correlated items as a cohesive unit rather than independent entities.

Based on the above intuitions, we propose a graph-based partitioning strategy as detailed in Algorithm~\ref{alg:item-placement}. \emph{First}, we address the popularity skew by replicating hot items (Lines 1–12). To eliminate cache misses for the most frequent queries, we identify the top 0.1\% most popular items based on historical access frequency and replicate their KV caches across all model instances. Because the absolute number of these hot items is small, this replication strategy incurs negligible memory overhead while guaranteeing local access for the most common requests.
\emph{Second}, we construct an item similarity graph to manage the placement of the rest long-tail items (Lines 13–22). The nodes in the graph represent items and edge weights represent relevance derived from co-occurrence in historical requests. We then apply METIS~\cite{karypis1998multilevelk} to partition this graph into $k$ subgraphs ($k$ is the number of instances). By minimizing the edge cuts between partitions, this step ensures that strongly correlated items are grouped onto the same physical node, thereby maximizing cache affinity for multi-item requests while maintaining balanced memory usage across the cluster.
The offline placement is not static. To accommodate catalog evolution and popularity drifts, we periodically refresh the placement. Structural updates (e.g., new item ingestion) or significant distributional shifts trigger a background re-execution of Algorithm~\ref{alg:item-placement}. 

\begin{algorithm}[t]\scriptsize
\caption{Similarity-aware item placement algorithm with global replicas}
\label{alg:item-placement}
\begin{algorithmic}[1]
  \STATE \textbf{Input}: items $I$, reviews $R$, number of instances $k$
  \STATE \textbf{Output}: $k$ subgraphs $\{G_1, \dots, G_k\}$

  \STATE $\triangleright$ Phase 1: Compute item popularity
  \FOR{each item $i \in I$}
    \STATE $h_i = |\{r \in R : r.\text{item\_id} = i\}|$
  \ENDFOR

  \STATE Sort items in descending order of $h_i$
  \STATE Let $I_{\text{hot}}$ be the top $0.1\%$ items
  \STATE Let $I_{\text{cold}} = I \setminus I_{\text{hot}}$

  \STATE Initialize node set $V = \emptyset$ and heat map $H$

  \STATE $\triangleright$ Phase 2: Create global replicas for hot items
  \FOR{each item $i \in I_{\text{hot}}$}
    \FOR{$j = 1$ to $k$}
      \STATE Create replica $i^{(j)}$ for instance $j$
      \STATE Add $i^{(j)}$ to $V$; set $H[i^{(j)}] = h_i / k$
    \ENDFOR
  \ENDFOR

  \STATE $\triangleright$ Phase 3: Add cold items
  \FOR{each item $i \in I_{\text{cold}}$}
    \STATE Add $i$ to $V$; set $H[i] = h_i$
  \ENDFOR

  \STATE $\triangleright$ Phase 4: Build item similarity graph
  \STATE $G \leftarrow (V, \emptyset)$
  \STATE $E \leftarrow \textsc{BuildItemSimilarityGraph}(V, I, R)$
  \STATE $G \leftarrow (V, E)$

  \STATE $\triangleright$ Phase 5: Graph partitioning
  \STATE $\{G_1, \dots, G_k\} \leftarrow \textsc{PartGraphByMetis}(G, k)$

  \STATE \textbf{return} $\{G_1, \dots, G_k\}$
\end{algorithmic}
\end{algorithm}



\subsection{Online Phase: Cache-Aware Distributed Inference}
\label{sec:online-phase}

The online phase executes the stateful inference pipeline introduced in the overview (\S\ref{sec:design-overview}). It orchestrates three pipelined stages to minimize TTFT while preserving model fidelity, including \emph{affinity scheduling}, \emph{zero-copy retrieval}, and \emph{selective recomputation}.

\subsubsection{\bf Cache-Aware Global Scheduling}

Standard load balancers (e.g., Round-Robin or Least-Loaded~\cite{theaibrixteam2025aibrixscalablecosteffectivelarge}) are agnostic to the distributed state, leading to frequent cache misses in our sharded architecture. To maximize the utility of distributed KV cache, \sysname{} employs a \emph{global} scheduler that synchronizes item placement data with real-time node utilization.

The scheduler routes incoming requests by optimizing two competing goals: \emph{data locality} (minimize item retrieval overhead) and \emph{load balancing} (prevent hotspots on popular shards). We quantify this trade-off using a request-node \emph{affinity score}:$$\text{Affinity}(R, p) = \alpha \cdot \widehat{\text{Hit}}(R, p) + \beta \cdot (1 - \text{Load}(p)) \quad (2)$$Here, $\widehat{\text{Hit}}(R, p)$ estimates the item-cache hit ratio for request $R$ on instance $p$ (e.g., $|I(R) \cap C(p)| / |I(R)|$, where $I(R)$ is the set of requested candidate items and $C(p)$ is the set of items cached on $p$). $\text{Load}(p)$ denotes the normalized runtime load (e.g., GPU utilization or queue depth). Hyperparameters $\alpha$ and $\beta$ allow the system to dynamically adapt between cache-priority and load-priority depending on traffic intensity. We empirically study the impact of the two parameters in \S\ref{sec:ablation-scheduling}.

\subsubsection{\bf Zero-Copy Retrieval and Selective Attention}

Once a request reaches the target instance, the local KV cache manager constructs the prompt context. This module addresses the challenge of assembling non-contiguous semantic blocks while correcting the approximation errors inherent in reuse.

\emph{a) KV Block Retrieval.} The prompt is decomposed into three components, each handled differently to ensure correctness: i) \emph{Instruction tokens}, which are instance-dependent and require strict adherence to system prompts. They are always fully recomputed.
ii) \emph{Review tokens}, which exhibit high semantic redundancy and are mapped to the nearest pre-computed semantic prototype using LSH-based matching~\cite{jafari2021survey}. 
Crucially, to prevent incorrect cross-context contamination, caching is strictly limited to natural language content, while instance-specific fields (e.g., timestamps, item identifiers, and delimiters) are isolated and always recomputed.
iii) \emph{Candidate item tokens}, which represent immutable catalog data and are retrieved from the local CPU cache using Item IDs. Valid hits are transferred directly to GPU memory via a zero-copy path~\cite{kato2013zerocopy}, while cache misses are computed on-the-fly.

\emph{b) Selective Recomputation.} 
To mitigate the semantic bias (for reviews) or contextual misalignment (for items) introduced by stitching cached blocks, \sysname{} implements a system-level compensation mechanism. 

Leveraging the observation that token importance persists across layers~\cite{zhao2025semsharekv,yao2025cacheblend}, we perform full attention computation in the first decoder layer to identify ``heavy hitters'' (i.e., structurally critical tokens requiring update). 
We select these tokens using an \emph{importance score} $S_i$:
$$S_i = (1 - \lambda)\|A_i\|_1 + \lambda \sum_{M \in \{K,V\}} \|M_i^{\text{new}} - M_i^{\text{cached}}\|_1 \quad (3)$$where $A_i$ is the attention weight of token $i$. The second term measures the divergence between the current context and the cached state. For review tokens, $\lambda$ balances structural importance with semantic drift. For item tokens, where the cached block is a lexical match, the divergence term vanishes, and selection relies solely on $A_i$.

In subsequent layers, \sysname{} recomputes exact attention only for these heavy hitters and the local sliding window, reusing cached states for the remainder of the sequence.

\subsubsection{\bf End-to-End Prefill Pipeline}

The final execution pipeline integrates these steps to minimize the critical path: (i) \emph{Assembly}: KV blocks are gathered from the replicated library and local item shard; (ii) \emph{Alignment}: Reused blocks undergo positional adjustment (e.g., RoPE rotation~\cite{su2024roformer}) to align with the current request's indices; (iii) \emph{Correction}: The selective recomputation kernel updates heavy hitters. We optimize this pipeline by overlapping the first-layer attention computation with the PCI-e transfer of KV blocks from CPU to GPU, effectively hiding the transmission latency.

\subsection{System Implementation}

To validate RcLLM’s design across both performance and accuracy dimensions, we developed a hybrid implementation framework comprising two distinct artifacts:
\begin{itemize}[leftmargin=*]
    \item {\bf Distributed Serving Engine}: We implemented RcLLM’s distributed logic on top of Vidur~\cite{MLSYS2024_b74a8de4}, a high-fidelity LLM inference simulator. We extended Vidur’s kernel to support our graph-based item placement, affinity-aware global scheduler, and non-contiguous KV retrieval. This implementation models the full distributed stack, including PCIe transfers, 100Gbps network interconnects, and request queueing dynamics, enabling accurate cluster-scale latency measurements.
    \item {\bf Accuracy Prototype}: For verifying recommendation fidelity, we built a functional prototype using HuggingFace transformers~\cite{wolf2020transformers}. We modified the underlying attention mechanism of Qwen3-8B to support selective attention, allowing us to inject sparse attention masks and measure the exact ranking impact of our semantic and item KV reuse strategies on real datasets.
\end{itemize}

\section{Evaluation}
\label{sec:evaluation}

\begin{figure}[t]
  \centering
  \begin{subfigure}[b]{0.32\columnwidth}
    \centering
    \includegraphics[width=1.05\textwidth]{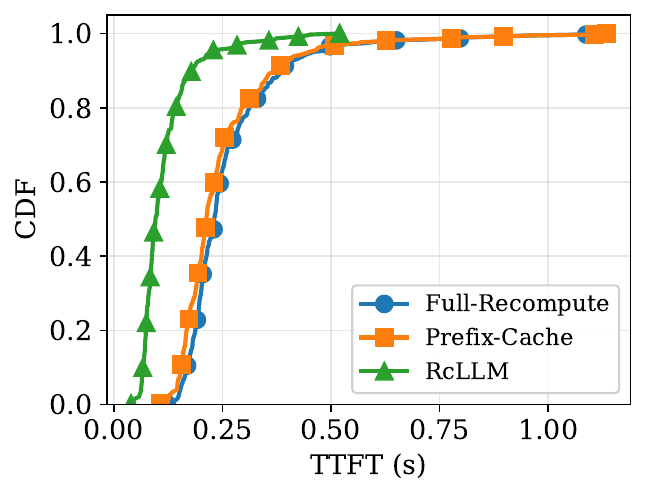}\vspace{-1ex}
    \caption{\scriptsize Amazon (Qwen3-8B)}
    \label{fig:amazon-latency-8b}
  \end{subfigure}
  \hfill
  \begin{subfigure}[b]{0.32\columnwidth}
    \centering
    \includegraphics[width=1.05\textwidth]{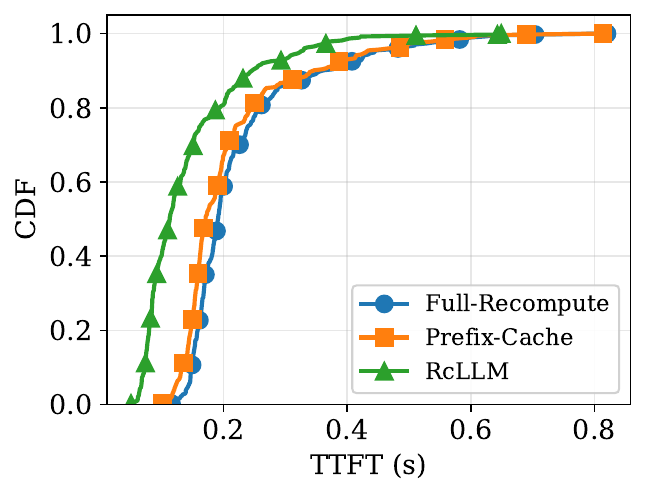}\vspace{-1ex}
    \caption{\scriptsize Yelp (Qwen3-8B)}
    \label{fig:yelp-latency-8b}
  \end{subfigure}
  \hfill
  \begin{subfigure}[b]{0.32\columnwidth}
    \centering
    \includegraphics[width=1.05\textwidth]{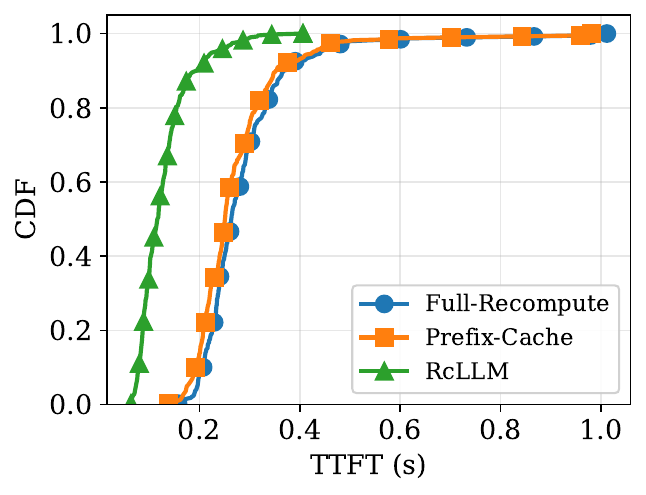}\vspace{-1ex}
    \caption{\scriptsize Goodreads (Qwen3-8B)}
    \label{fig:goodread-latency-8b}
  \end{subfigure}

  \vspace{2ex} 

  \begin{subfigure}[b]{0.32\columnwidth}
    \centering
    \includegraphics[width=1.05\textwidth]{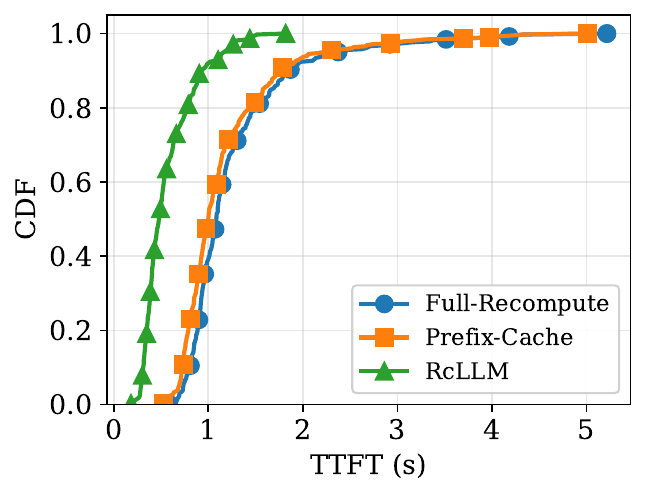}\vspace{-1ex}
    \caption{\scriptsize Amazon (Qwen-72B)}
    \label{fig:amazon-latency-72b}
  \end{subfigure}
  \hfill
  \begin{subfigure}[b]{0.32\columnwidth}
    \centering
    \includegraphics[width=1.05\textwidth]{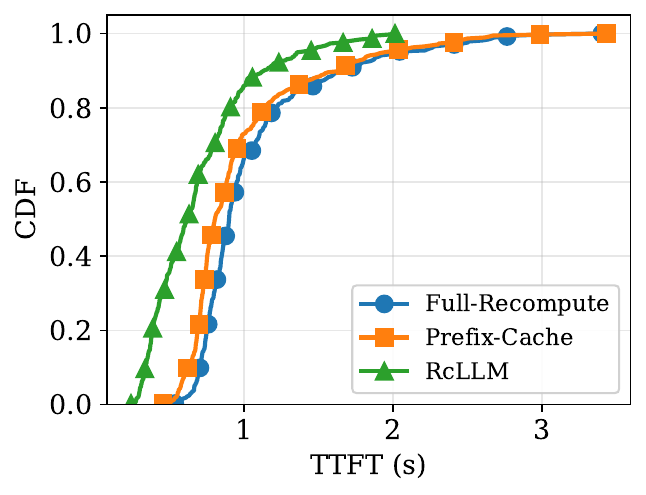}\vspace{-1ex}
    \caption{\scriptsize Yelp (Qwen-72B)}
    \label{fig:yelp-latency-72b}
  \end{subfigure}
  \hfill
  \begin{subfigure}[b]{0.32\columnwidth}
    \centering
    \includegraphics[width=1.05\textwidth]{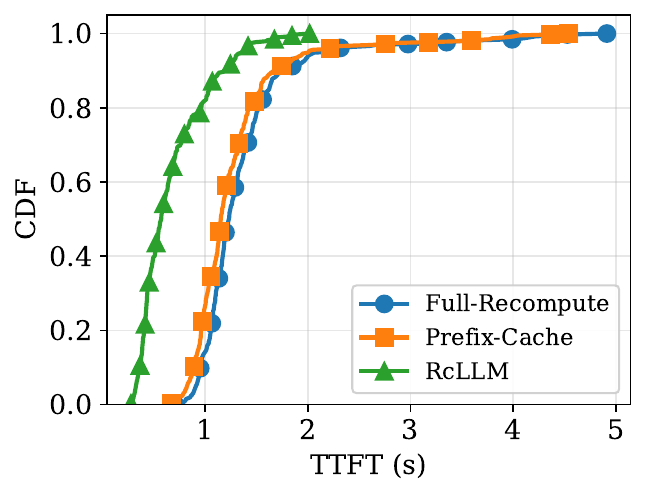}\vspace{-1ex}
    \caption{\scriptsize Goodreads (Qwen-72B)}
    \label{fig:goodread-latency-72b}
  \end{subfigure}
  
  \vspace{-1ex}
  \caption{TTFT CDF comparison in a distributed setting with $K{=}40$ instances for Qwen3-8B (top) and Qwen-72B (bottom) across three datasets. \sysname{} consistently shifts the distribution left, demonstrating significant latency reductions.}
  \label{fig:latency-cdf-combined}
\end{figure}

\begin{figure*}[t]
  \centering
  \begin{subfigure}[b]{0.48\columnwidth}
    \centering
    \includegraphics[width=1.1\textwidth]{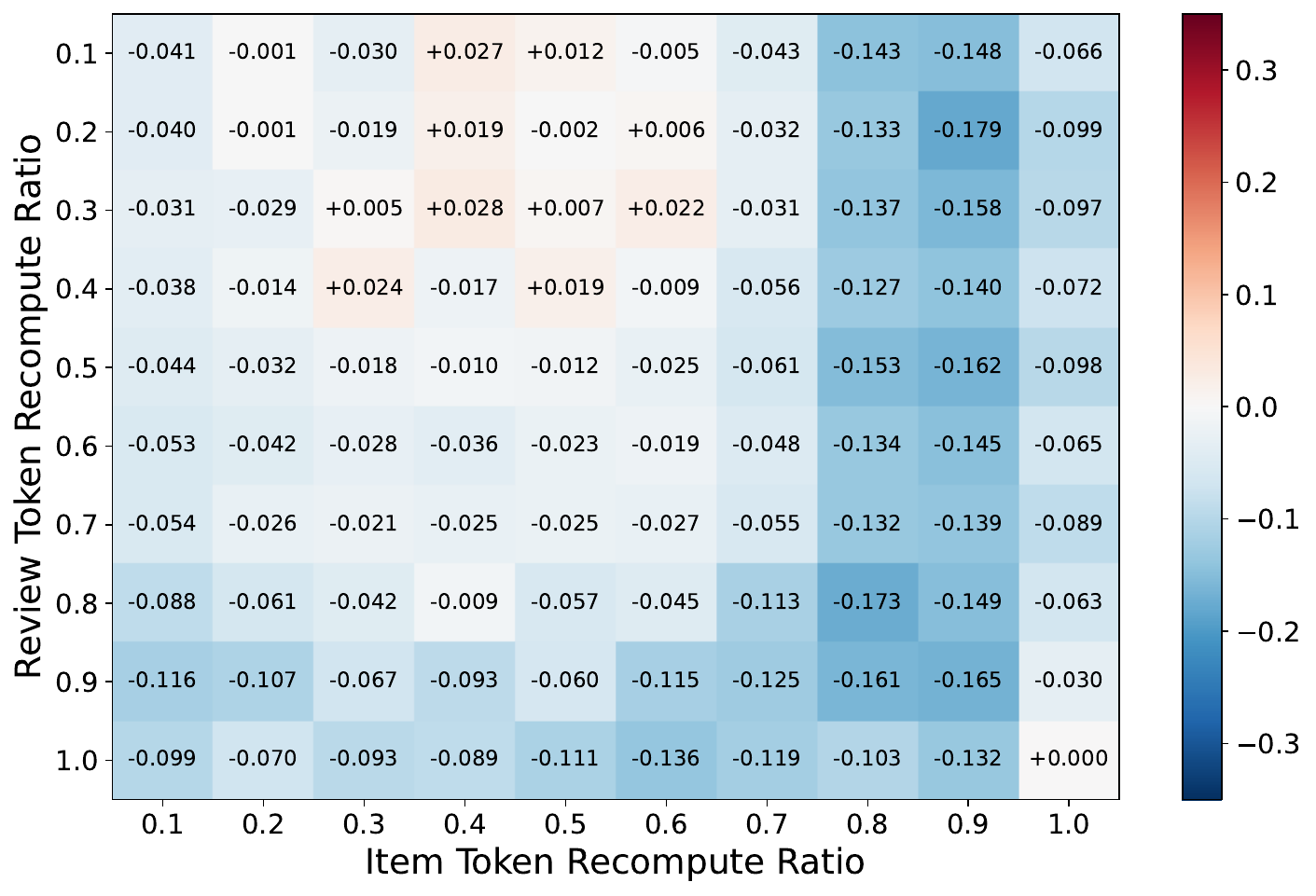}    \vspace{-3ex}
    \caption{\scriptsize\bf Amazon}
    \label{fig:amazon-ratio-analysis}
  \end{subfigure}
  \hfil
  \begin{subfigure}[b]{0.48\columnwidth}
    \centering
    \includegraphics[width=1.1\textwidth]{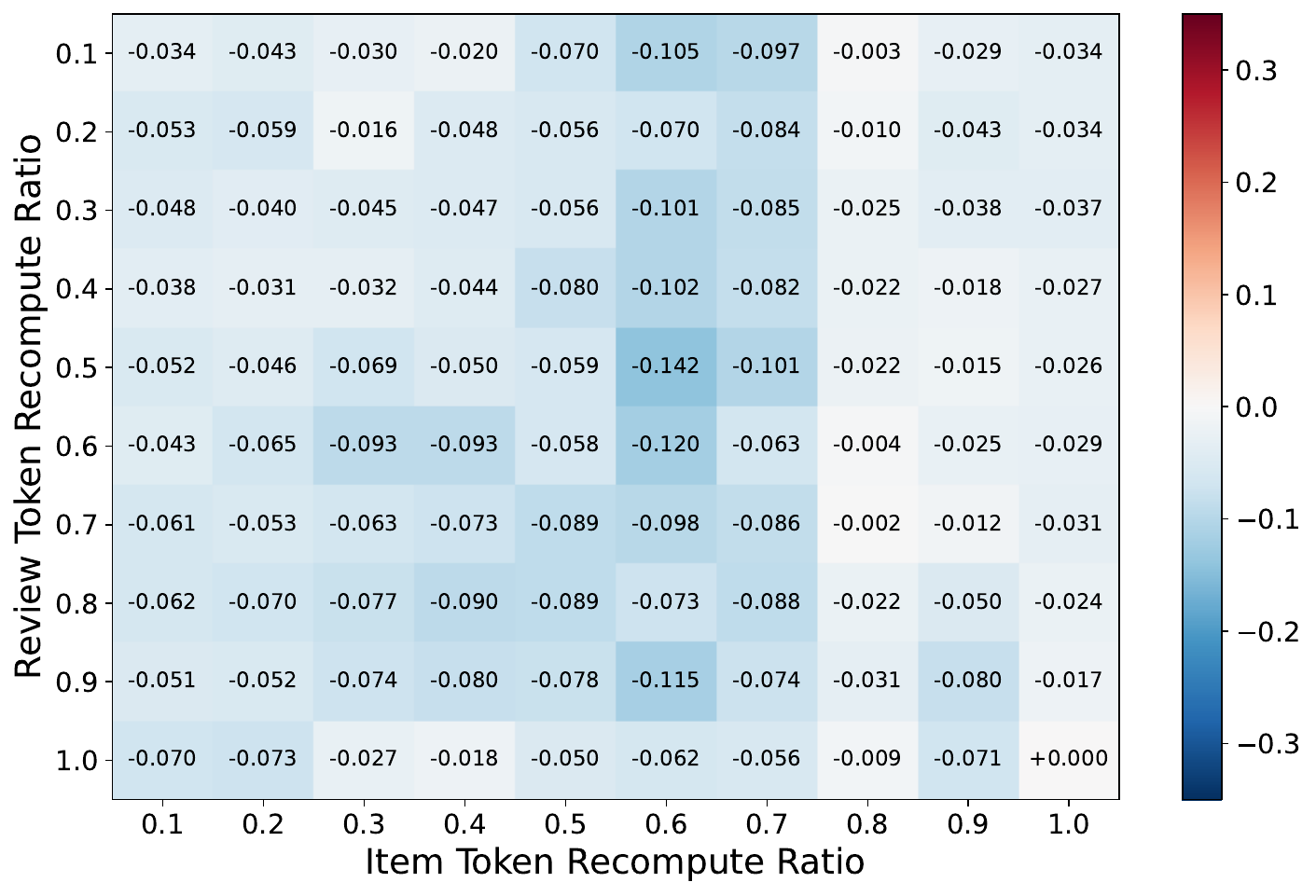}    \vspace{-3ex}
    \caption{\scriptsize\bf Goodreads}
    \label{fig:goodreads-ratio-analysis}
  \end{subfigure}
  \hfil
  \begin{subfigure}[b]{0.48\columnwidth}
    \centering
    \includegraphics[width=1.1\textwidth]{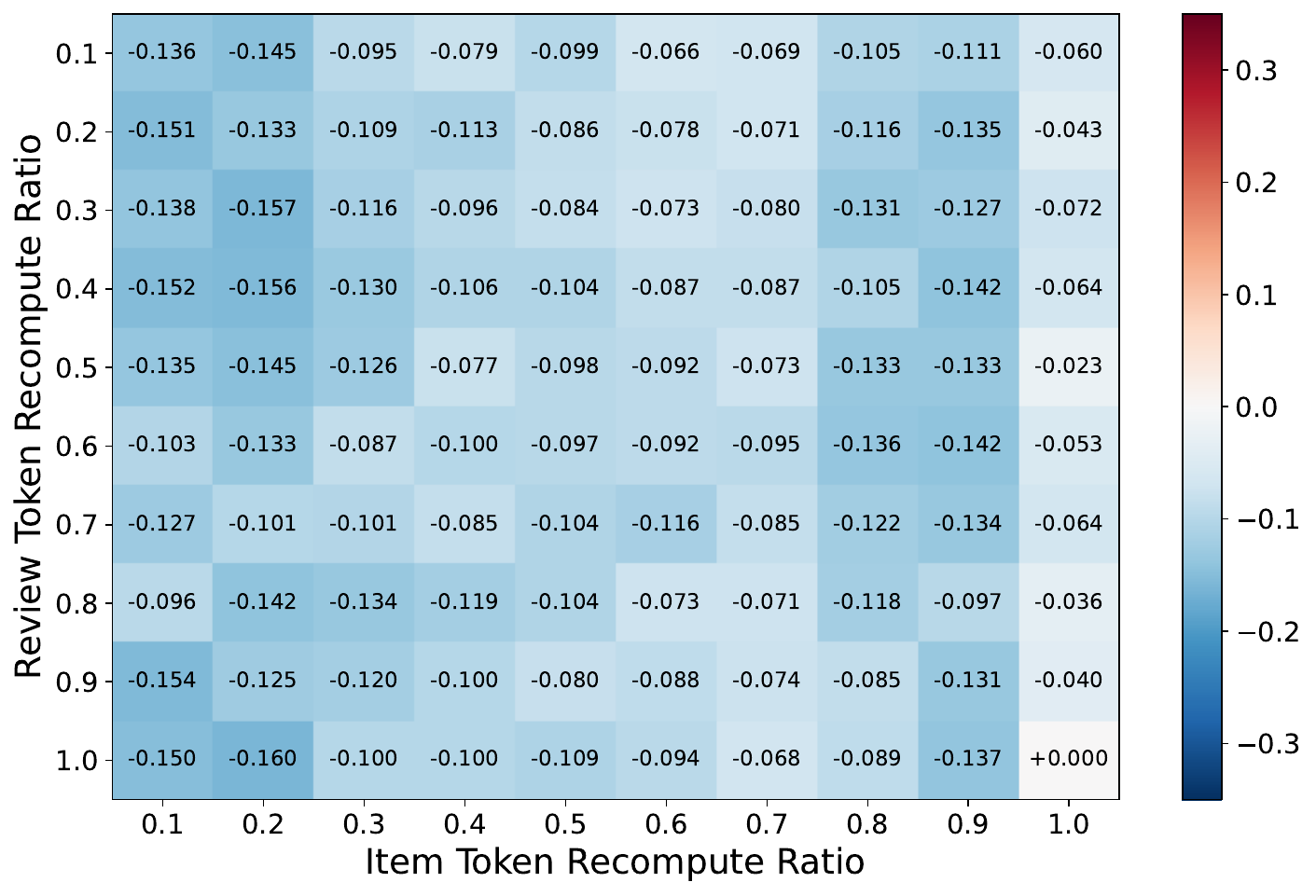}    \vspace{-3ex}
    \caption{\scriptsize\bf Yelp}
    \label{fig:yelp-ratio-analysis}
  \end{subfigure}
  \vspace{-1ex}
  \caption{Difference between the NDCG of \sysname{} and that of Full-Recompute (the higher the better).
  \sysname{} maintains ranking quality under moderate recomputation budgets,
  enabling substantial KV reuse with limited accuracy loss.}
  \label{fig:ratio-analysis}
\end{figure*}

\renewcommand{\arraystretch}{0.9}
\begin{table*}[t]\footnotesize
\centering
\caption{Recommendation accuracy comparison. Best result is bolded and second best result is underlined.}
\label{tab:accuracy_comparison}
\begin{tabular}{llcccccccc}
\toprule
\textbf{Dataset} & \textbf{Method} & \textbf{HR@1} & \textbf{HR@3} & \textbf{HR@5} & \textbf{HR@10} & \textbf{MRR} & \textbf{NDCG@5} & \textbf{NDCG@10} & \textbf{NDCG@20}\\
\midrule
\multirow{4}{*}{Amazon}
& Full-Recompute & \underline{0.160} & \textbf{0.380} & \textbf{0.520} & \underline{0.700} & \underline{0.333} & \underline{0.346} & \underline{0.403} & \underline{0.480}\\
& \sysname{} ($r_{\text{item}}{=}0.3, r_{\text{rev}}{=}0.3$) & \textbf{0.220} & \underline{0.340} & \underline{0.500} & \textbf{0.740} & \textbf{0.360} & \textbf{0.360} & \textbf{0.438} & \textbf{0.485}\\
& CacheBlend & 0.142 & 0.300 & 0.420 & 0.620 & 0.294 & 0.295 & 0.360 & 0.403\\
& EPIC & 0.100 & 0.200 & 0.280 & 0.440 & 0.209 & 0.191 & 0.241 & 0.346\\
\midrule
\multirow{4}{*}{Goodreads}
& Full-Recompute & \textbf{0.163} & \textbf{0.408} & \textbf{0.510} & \textbf{0.776} & \textbf{0.338} & \textbf{0.344} & \textbf{0.427} & \textbf{0.484}\\
& \sysname{} ($r_{\text{item}}{=}0.3, r_{\text{rev}}{=}0.2$) & \underline{0.143} & \underline{0.286} & \underline{0.469} & \textbf{0.776} & \underline{0.287} & \underline{0.295} & \underline{0.393} & \underline{0.468}\\
& CacheBlend & 0.122 & 0.224 & 0.387 & 0.653 & 0.244 & 0.246 & 0.329 & 0.370\\
& EPIC & 0.102 & 0.265 & 0.387 & 0.551 & 0.249 & 0.248 & 0.300 & 0.400\\
\midrule
\multirow{4}{*}{Yelp}
& Full-Recompute & \textbf{0.256} & \textbf{0.558} & \textbf{0.628} & \underline{0.698} & \textbf{0.413} & \textbf{0.456} & \textbf{0.477} & \textbf{0.497}\\
& \sysname{} ($r_{\text{item}}{=}0.4, r_{\text{rev}}{=}0.5$) & \underline{0.140} & \underline{0.442} & \underline{0.535} & \textbf{0.744} & \underline{0.307} & \underline{0.341} & \underline{0.407} & \underline{0.414}\\
& CacheBlend & 0.043 & 0.128 & 0.277 & 0.489 & 0.150 & 0.149 & 0.220 & 0.255\\
& EPIC & 0.021 & 0.192 & 0.319 & 0.468 & 0.153 & 0.169 & 0.219 & 0.259\\
\bottomrule
\end{tabular}
\end{table*}


We evaluate RcLLM along three dimensions: (1) \emph{End-to-end Latency}, quantifying speedups in distributed serving; (2) \emph{Recommendation Accuracy}, assessing the fidelity of selective recomputation; and (3) \emph{System Ablations}, isolating the benefits of placement and scheduling.

\subsection{Experimental Setup}
\label{sec:eval-config}

\textbf{Testbed Environment.}
We simulate a local cluster with $K=40$ instances by default and scale to $K\le100$ in scalability ablations. For experiments with Qwen3-8B~\cite{yang2025qwen3technicalreport}, each instance is equipped with a single NVIDIA A100 GPU (80GB). For the larger Qwen-72B~\cite{yang2024qwen2technicalreport}, each instance utilizes 4 NVIDIA A100 GPUs interconnected via pairwise NVLink, with the model deployed using Tensor Parallelism (TP=4) to meet its memory and computational demands. All instances are connected via a 100Gbps high-bandwidth interconnect.

We employ Qwen3-8B as the primary model for evaluating recommendation accuracy, while Qwen-72B is additionally used to assess system performance and scalability. This selection reflects a balance between reasoning capability and industrial serving costs. Models with even larger parameters are not considered due to the prohibitive gap between their baseline inference latency and the stringent recommendation SLOs (Figure~\ref{fig:latency-test}).

\textbf{Datasets and Workloads.} We evaluate on three widely used recommendation datasets, using raw review data to synthesize prompts following the structure in \S\ref{sec:motivation} (i.e., Instruction + History + Items):
\begin{itemize}[leftmargin=*]
    \item Amazon Reviews~\cite{hou2024bridginglanguageitemsretrieval}: Covers three categories (Industrial, Musical Instruments, and Video Games) to balance domain diversity.
    \item Yelp~\cite{yelp2025} \& Goodreads~\cite{wan2018item}: Selected to stress-test performance on datasets with longer, verbose user histories. For Goodreads, we include three categories: Children, Graphic, and Poetry.
    \item Trace Generation: Online inference traffic is generated using real-world interaction traces sourced from~\cite{yan2025agentsociety}.
\end{itemize}

\textbf{Baselines.}
We compare \sysname{} against two categories of baselines, namely {system baselines} for latency comparison and {algorithmic baselines} for accuracy comparison.
\begin{itemize}[leftmargin=*]
    \item \emph{System Baselines}: 1) Full-Recompute, which is the standard autoregressive inference with no reuse; 2) Prefix-Cache (SOTA), which is the industrial standard (e.g., vLLM~\cite{vllm2023}, SGLang~\cite{zheng2024sglang}) reusing only contiguous prefixes.
    \item \emph{Algorithmic Baselines}: 1) CacheBlend~\cite{yao2025cacheblend}, a beyond-prefix KV reuse baseline that caches KV states for reusable context chunks and blends cached knowledge with the current prompt to approximate full attention context (widely used in RAG-style serving); 2) EPIC~\cite{hu2025epic}, a position-independent KV caching baseline that decouples/corrects positional effects so cached KV blocks can be reused even when inserted at different positions, enabling non-prefix reuse across requests.
\end{itemize}
Note that CacheBlend and EPIC lack distributed scheduling logic, so they are compared only on accuracy benchmarks.

\textbf{Metrics.}
For latency evaluation, we compare Time-To-First-Token (TTFT) at P50, P90, and P99. TTFT is the bottleneck metric for generative recommendation as systems act immediately on the first emitted item. For accuracy evaluation, we adopt a few metrics commonly used in recommendation systems, including Hit Rate (HR@K), Mean Reciprocal Rank (MRR), and NDCG@K.

\textbf{Configurations.}
We control approximation aggressiveness via recompute ratios ($r_{rev}$ for history, $r_{item}$ for items). Based on the sensitivity analysis in \S\ref{sec:eval-accuracy} and \S\ref{sec:ablation-recompute}, we use the following defaults which maximize speed while maintaining accuracy parity: $r_{rev}$=$r_{item}$=0.3 for Amazon,  $r_{rev}$=0.2, $r_{item}$=0.3 for Goodreads and $r_{rev}$=0.5, $r_{item}$=0.4 for Yelp.

\subsection{End-to-End Latency Performance}\label{sec:eval-latency}

\begin{figure*}[t]
  \centering
  \begin{subfigure}[b]{0.48\columnwidth}
    \centering
    \includegraphics[width=\linewidth]{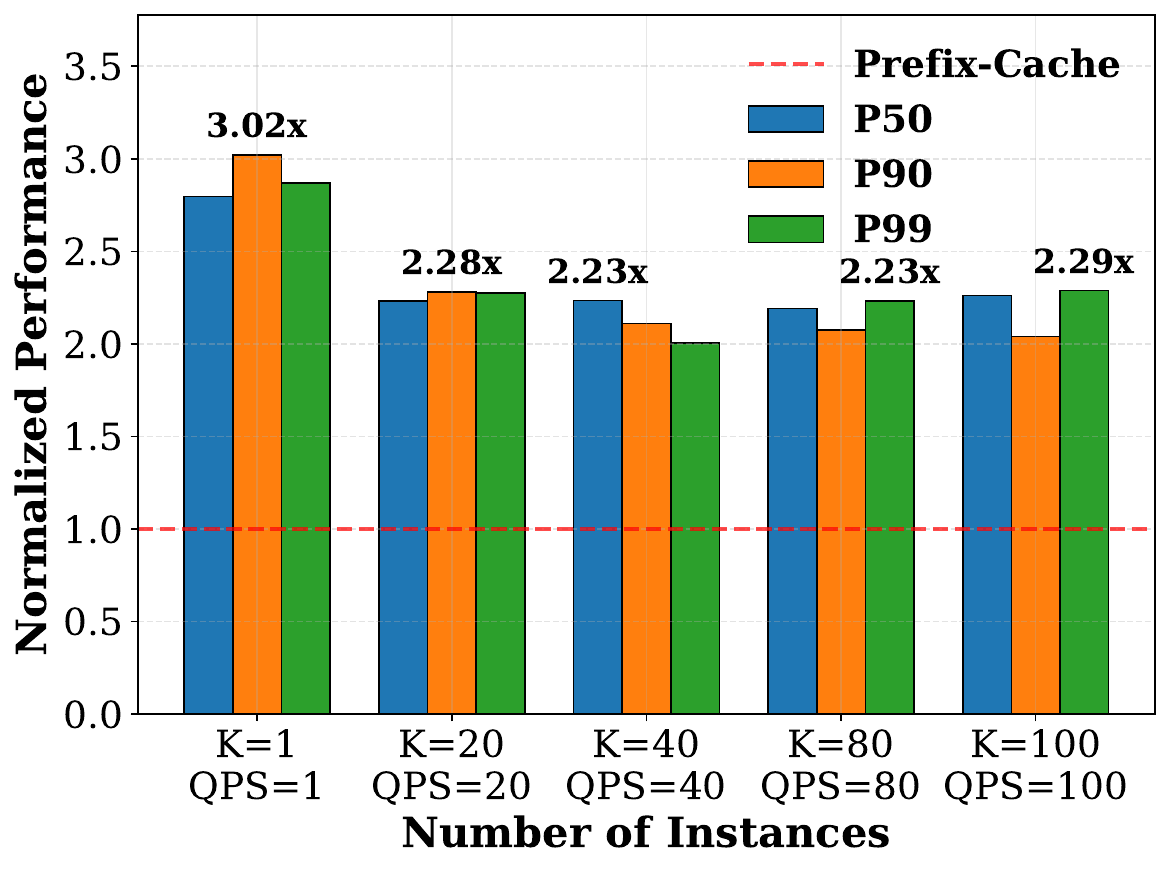}\vspace{-1ex}
    \caption{\scriptsize \textbf{Amazon (Qwen3-8B)}}
    \label{fig:amazon-speedup-8b}
  \end{subfigure}
  \hfil
  \begin{subfigure}[b]{0.48\columnwidth}
    \centering
    \includegraphics[width=\linewidth]{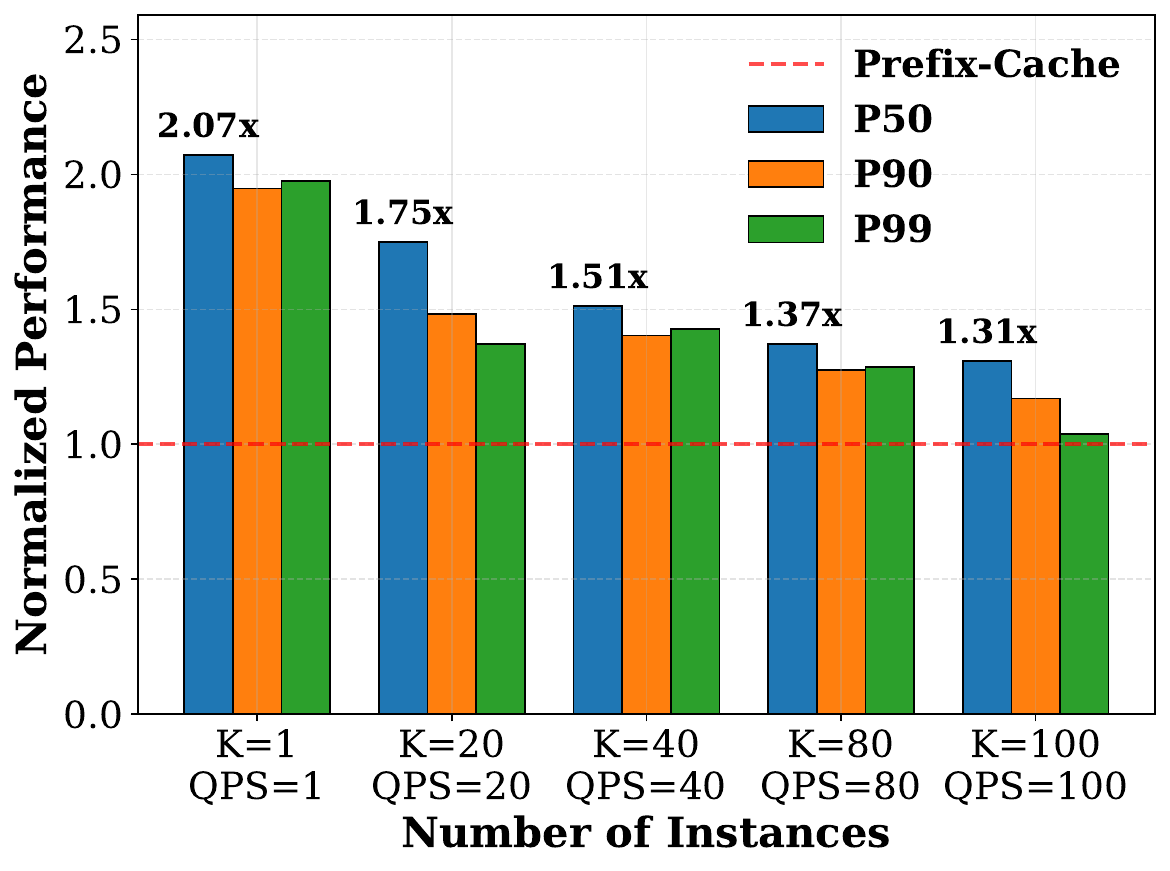}\vspace{-1ex}
    \caption{\scriptsize \textbf{Yelp (Qwen3-8B)}}
    \label{fig:yelp-speedup-8b}
  \end{subfigure}
  \hfil
  \begin{subfigure}[b]{0.48\columnwidth}
    \centering
    \includegraphics[width=\linewidth]{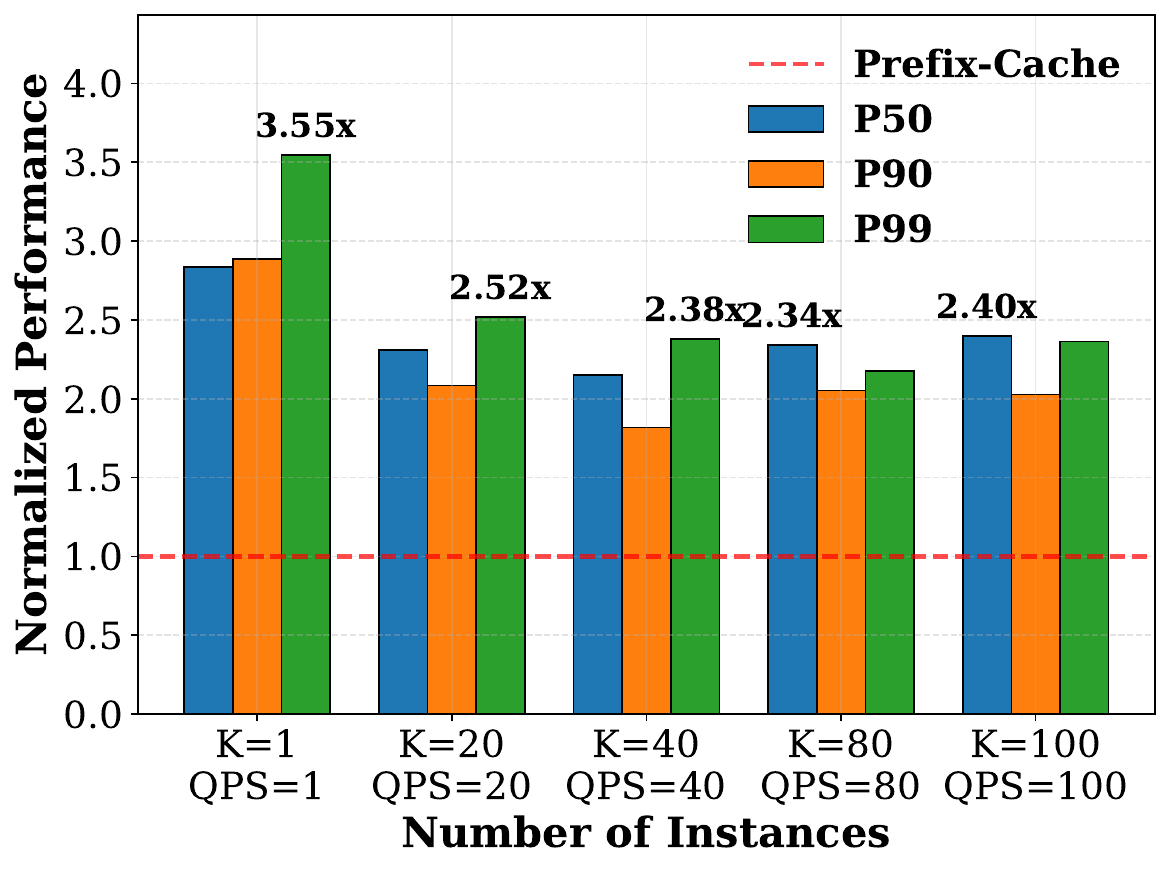}\vspace{-1ex}
    \caption{\scriptsize \textbf{Goodreads (Qwen3-8B)}}
    \label{fig:goodread-speedup-8b}
  \end{subfigure}

  \vspace{0.5ex}

  \begin{subfigure}[b]{0.48\columnwidth}
    \centering
    \includegraphics[width=\linewidth]{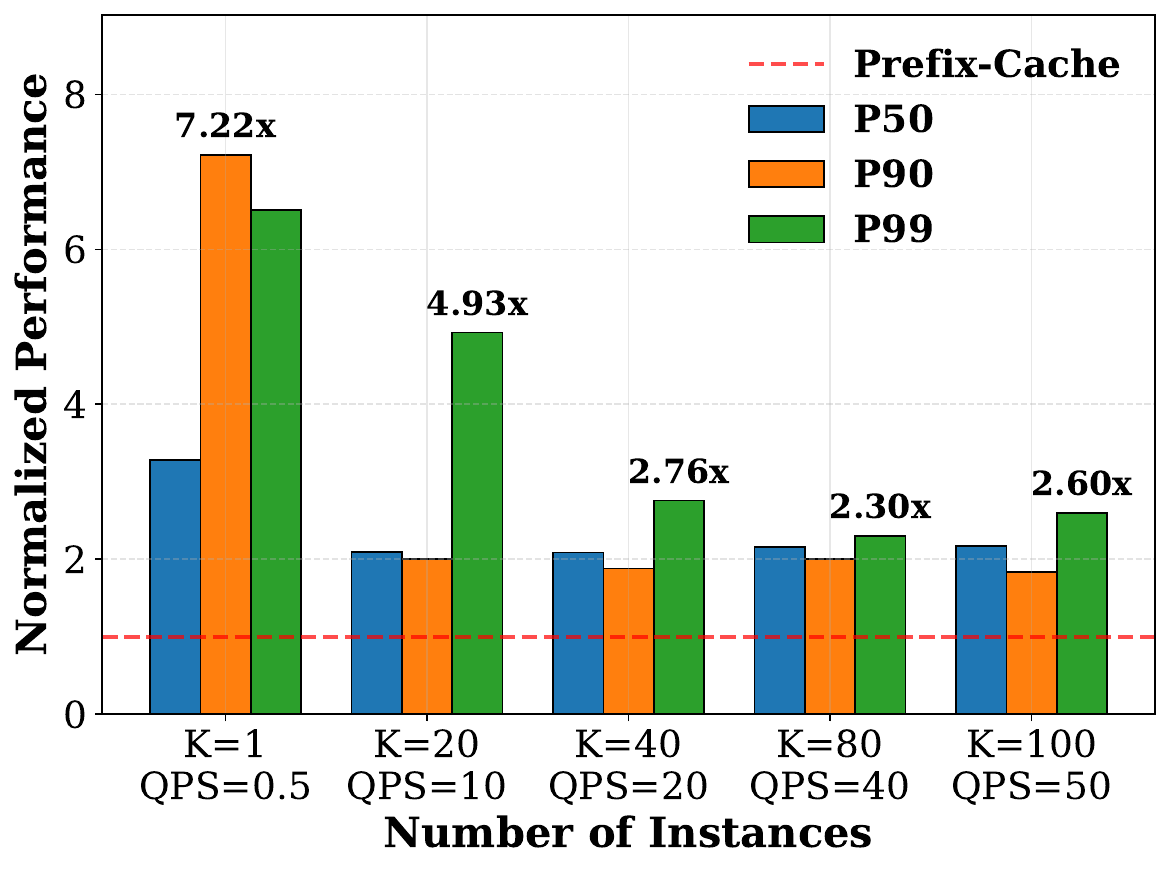}\vspace{-1ex}
    \caption{\scriptsize \textbf{Amazon (Qwen-72B)}}
    \label{fig:amazon-speedup-72b}
  \end{subfigure}
  \hfil
  \begin{subfigure}[b]{0.48\columnwidth}
    \centering
    \includegraphics[width=\linewidth]{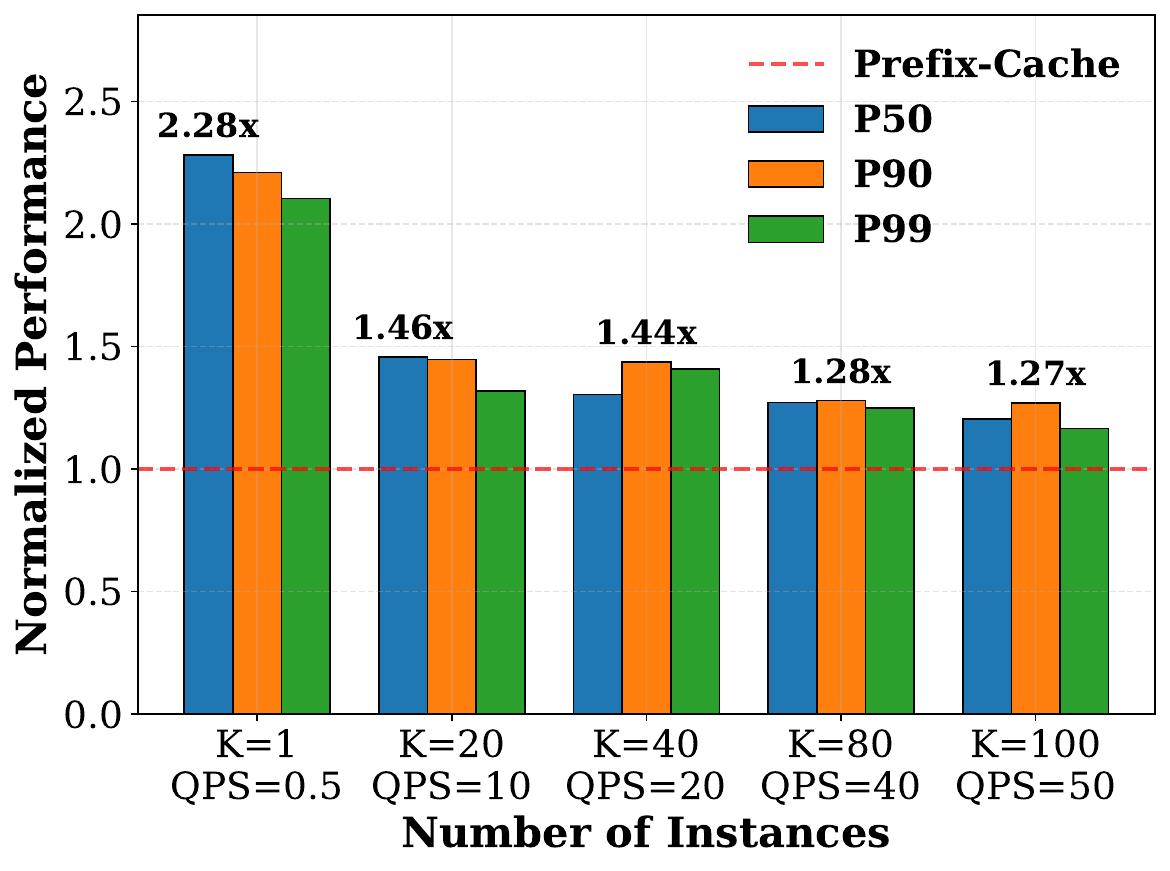}\vspace{-1ex}
    \caption{\scriptsize \textbf{Yelp (Qwen-72B)}}
    \label{fig:yelp-speedup-72b}
  \end{subfigure}
  \hfil
  \begin{subfigure}[b]{0.48\columnwidth}
    \centering
    \includegraphics[width=\linewidth]{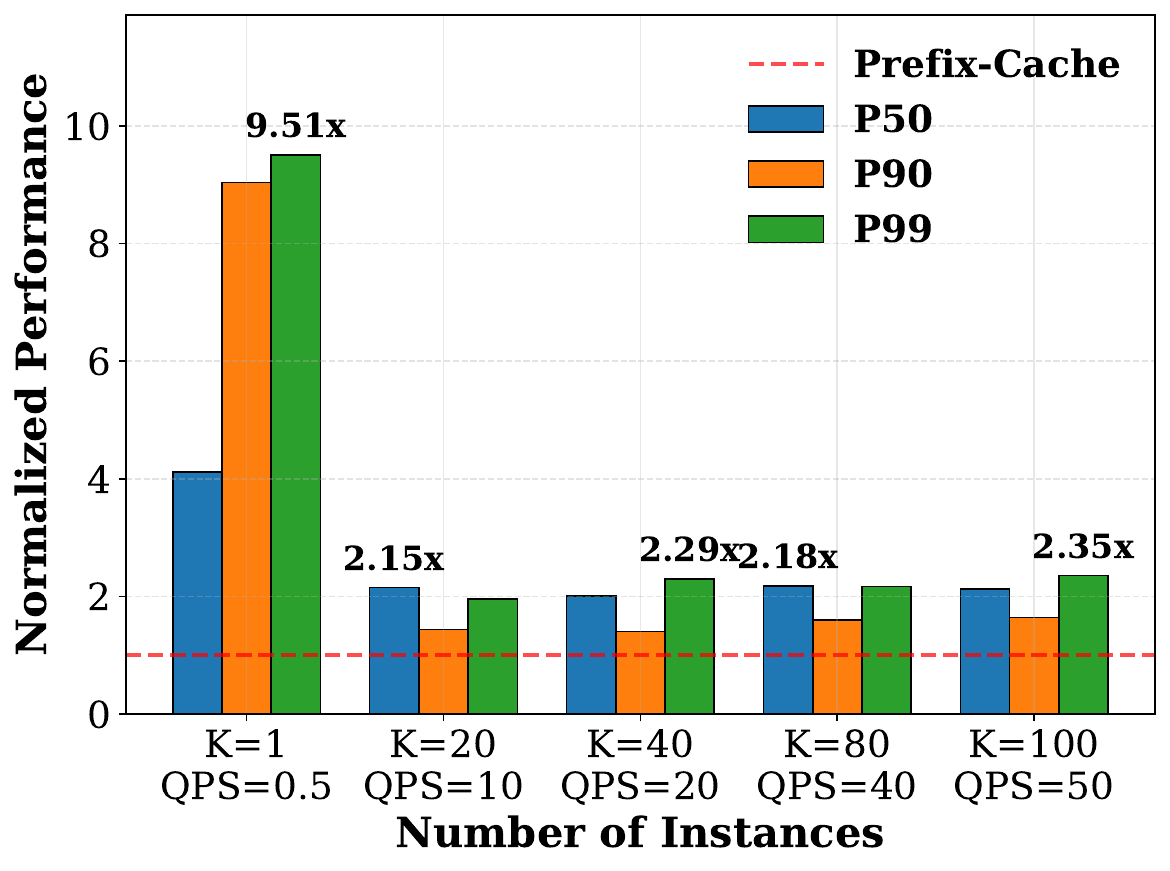}\vspace{-1ex}
    \caption{\scriptsize \textbf{Goodreads (Qwen-72B)}}
    \label{fig:goodread-speedup-72b}
  \end{subfigure}
  \vspace{-1ex}
  \caption{Normalized performance (speedup) of \sysname{} compared to Prefix-Cache under different cluster sizes for Qwen3-8B (top) and Qwen-72B (bottom). \sysname{} maintains significant speedups across both model scales and various cluster configurations.}
  \label{fig:instance-count}
\end{figure*}

\begin{figure}[t]
  \centering
  \begin{subfigure}[b]{0.45\linewidth}
    \centering
    \includegraphics[width=\linewidth]{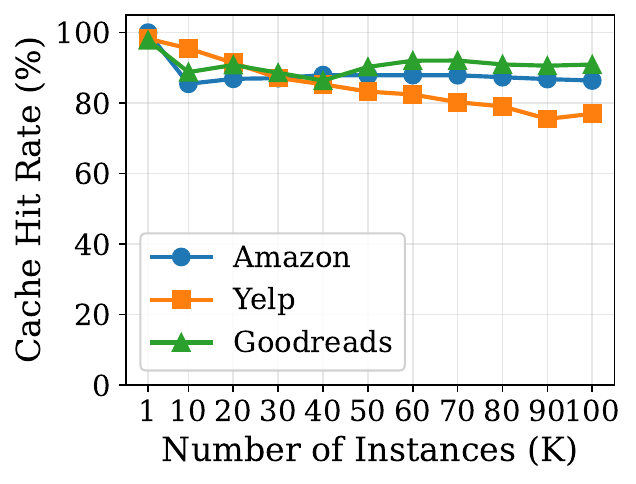}\vspace{-1ex}
    \caption{\scriptsize \textbf{Item-cache hit rate}}
    \label{fig:hit-rate}
  \end{subfigure}
  \hfil
  \begin{subfigure}[b]{0.45\linewidth}
    \centering
    \includegraphics[width=\linewidth]{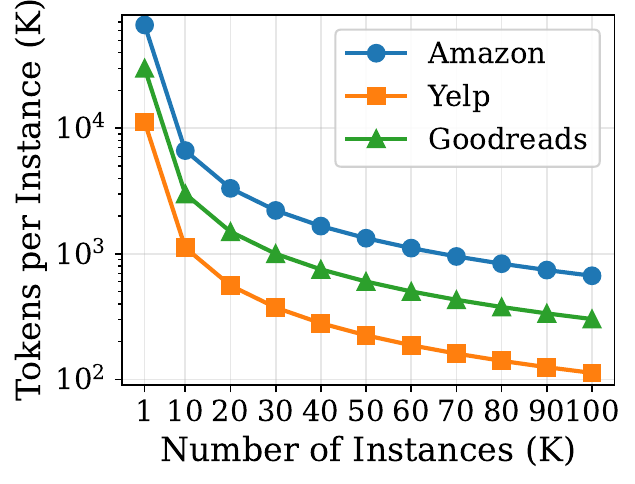}\vspace{-1ex}
    \caption{\scriptsize \textbf{Per-replica item-cache footprint}}
    \label{fig:item-storage}
  \end{subfigure}\vspace{-1ex}  
  \caption{The trade-off between locality and scalability under distributed sharding.
  (a) As $K$ increases, the item-cache hit rate naturally declines due to sharding; however, our similarity-aware placement maintains high locality (e.g., $\ge 75\%$ even at $K=100$).
  (b) The per-replica cached item footprint drops near-linearly with $K$, reducing the massive catalog to a manageable size for CPU caching.}
  \label{fig:hit-storage-instance-count}
\end{figure}

\begin{figure}[t]
  \centering
  \begin{subfigure}[b]{0.32\columnwidth}
    \centering
    \includegraphics[width=1.05\linewidth]{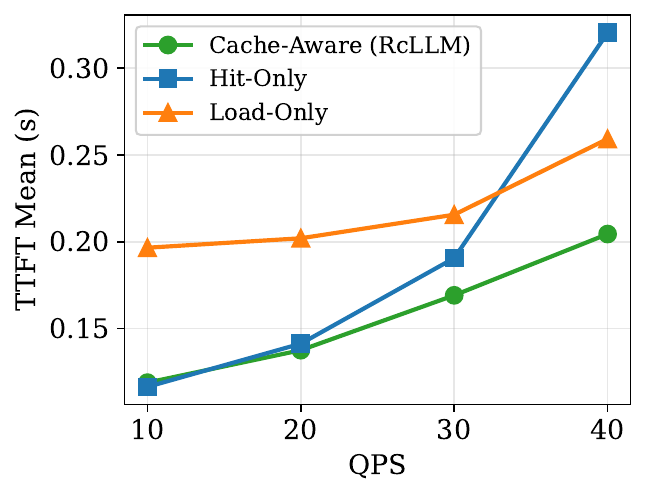}\vspace{-1ex}
    \caption{\scriptsize \textbf{Amazon}}
    \label{fig:amazon-scheduling}
  \end{subfigure}
  \hfil
  \begin{subfigure}[b]{0.32\columnwidth}
    \centering
    \includegraphics[width=1.05\linewidth]{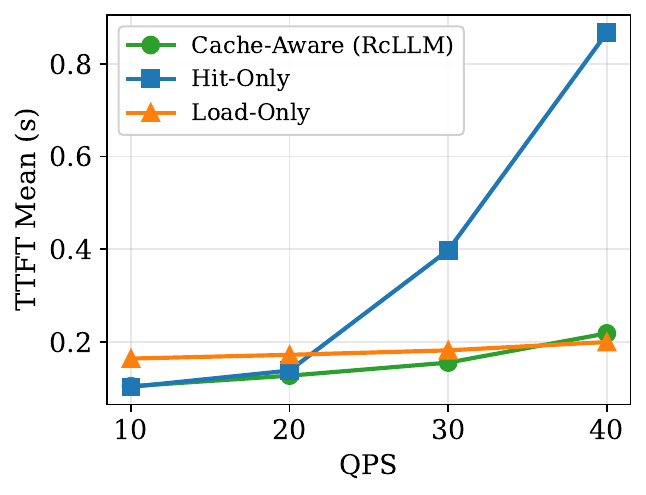}\vspace{-1ex}
    \caption{\scriptsize \textbf{Yelp}}
    \label{fig:yelp-scheduling}
  \end{subfigure}
  \hfil
  \begin{subfigure}[b]{0.32\columnwidth}
    \centering
    \includegraphics[width=1.05\linewidth]{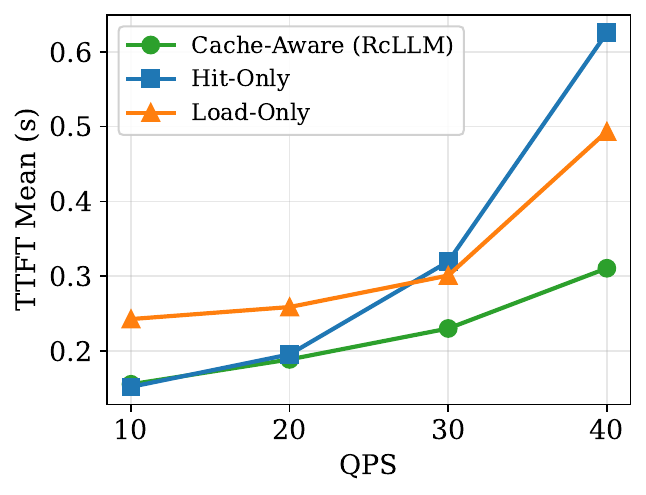}\vspace{-1ex}
    \caption{\scriptsize \textbf{Goodreads}}
    \label{fig:goodread-scheduling}
  \end{subfigure}
  \vspace{-1ex}
  \caption{The impact of scheduling policy on latency under increasing load.
  \sysname{} achieves the best (or near-best) mean TTFT by jointly optimizing cache locality and load balance.}
  \label{fig:scheduling}
\end{figure}

\begin{figure}[t]
  \centering
    \begin{subfigure}[t]{0.32\columnwidth} 
      \centering
      \includegraphics[width=1.05\linewidth]{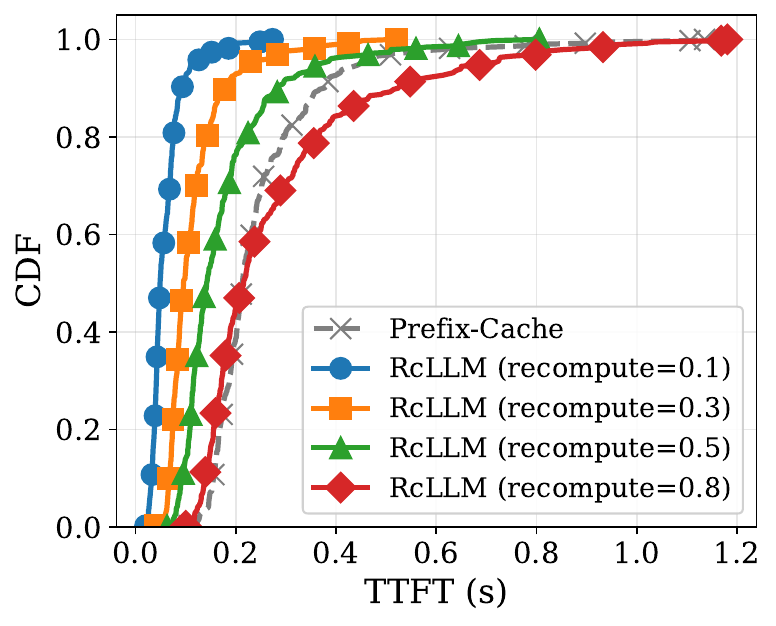}\vspace{-1ex}
      \caption{\scriptsize \textbf{Amazon}}
      \label{fig:amazon-recompute-8b}
    \end{subfigure}
    \hfill
    \begin{subfigure}[t]{0.32\columnwidth}
      \centering
      \includegraphics[width=1.05\linewidth]{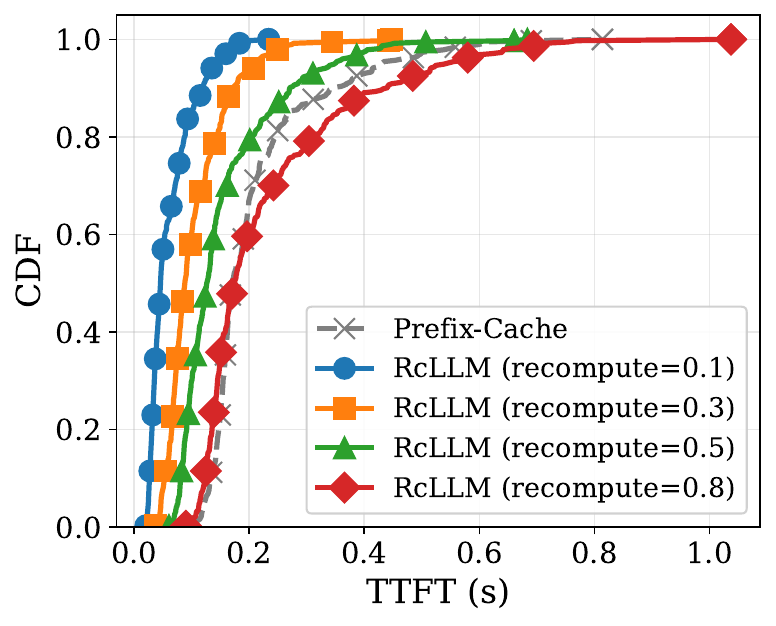}\vspace{-1ex}
      \caption{\scriptsize \textbf{Yelp}}
      \label{fig:yelp-recompute-8b}
    \end{subfigure}
    \hfill
    \begin{subfigure}[t]{0.32\columnwidth}
      \centering
      \includegraphics[width=1.05\linewidth]{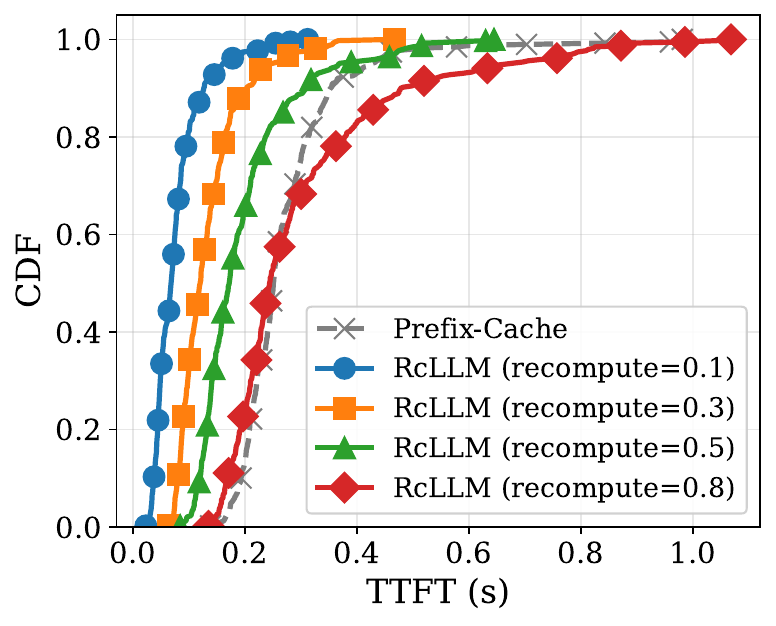}\vspace{-1ex}
      \caption{\scriptsize \textbf{Goodreads}}
      \label{fig:goodread-recompute-8b}
    \end{subfigure}
    \vspace{-1ex}
  \caption{The latency cost of increased fidelity ($r_{\text{rev}}{=}r_{\text{item}}{=}r$) for Qwen3-8B. Higher recomputation budgets shift the TTFT distributions right, illustrating the trade-off between computational cost and model fidelity.}
  \label{fig:recompute-latency}
\end{figure}


We focus our primary evaluation on the distributed setting ($K=40$ instances), which represents a realistic industrial deployment where the massive item catalog must be sharded. 
Figure~\ref{fig:latency-cdf-combined} shows the performance comparison results between \sysname{} and the two system baselines (i.e., Full-Recompute and Prefix-Cache).
We make several important observations.

\emph{First}, as shown in Figure~\ref{fig:latency-cdf-combined}, \sysname{} delivers substantial latency reductions across all datasets for both Qwen3-8B (top row) and Qwen-72B (bottom row). For Qwen3-8B, \sysname{} improves median (P50) TTFT by 1.5$\times$--2.2$\times$ and tail (P99) TTFT by 1.4$\times$--2.4$\times$; for Qwen-72B, it achieves 1.3$\times$--2.1$\times$ (P50) and 1.4$\times$--2.8$\times$ (P99) compared to Prefix-Cache, the industrial SOTA solution.
This effect is more pronounced for Qwen-72B, where a higher prefill cost amplifies the benefit of reuse.
This performance gap stems from the structural composition of recommendation prompts. Across our traces, the median prefill length spans 2.2K–3.0K tokens (with P90 reaching 4.0K–4.3K), yet the only truly shared contiguous prefix is the fixed system prompt (207 tokens), representing just $\sim$7–10\% of the prefill. Consequently, Prefix-Cache eliminates only a negligible fraction of the computation. In contrast, \sysname{} targets the dominant token mass: candidate items (66–82\% of prefill) and user histories (11–26\%). By decomposing the prompt and reusing these segments at the content-block granularity (item blocks and semantic prototypes), \sysname{} effectively bypasses the majority of the quadratic attention bottleneck that the baseline must fully recompute.
The gains are most pronounced on Amazon and Goodreads, reaching up to 2.8$\times$ speedup at P99, where the prefill is dominated by reusable item/history context.

\emph{Second}, beyond median improvements, \sysname{} significantly tightens the latency distribution for both model scales, avoiding the severe tail spikes seen in the baseline.
This stability is driven by the high local hit rates achieved by our placement and scheduling strategies. With similarity-aware placement and cache-aware routing, a single locality-optimal replica already covers a median of 90\% of a request’s candidate items at $K=40$ (mean 85\%--88\% across datasets).
Since item blocks are largely served locally from CPU memory and history KV reuse is local by construction (due to replication), only the residual item misses and sparse ``heavy-hitter'' tokens require online recomputation. This drastic reduction in critical-path work minimizes processing variance, yielding consistent improvements in both median and tail TTFT.


\subsection{Recommendation Accuracy}\label{sec:eval-accuracy}

We next evaluate whether \sysname{} can maintain ranking fidelity despite the approximations introduced by KV reuse. 
We compare against Full-Recompute (the ``Gold Standard'') and approximate baselines (CacheBlend, EPIC). Figure~\ref{fig:ratio-analysis} illustrates the trade-off between recomputation budget ($r$) and ranking accuracy. We make the following observations.

\emph{First}, we identify a stable ``Pareto frontier'' at moderate budgets where RcLLM matches Full-Recompute. On Amazon ($r=0.3$) and Goodreads ($r=[0.2,0.3]$), \sysname{} retains 96–100\% of the baseline NDCG while caching over 70\% of the prompt.
Surprisingly, on Amazon, \sysname{} slightly outperforms Full-Recompute on accuracy, boosting NDCG@10 from 0.403 to 0.438 (+8.6\%) and HR@10 from 0.70 to 0.74, as shown in Table~\ref{tab:accuracy_comparison}.
This counter-intuitive gain indicates that full recomputation often over-attends to noisy, template-heavy item metadata. By retrieving generic item prototypes and restricting high-precision compute to heavy hitter tokens via selective attention, \sysname{} effectively acts as a regularizer, filtering out semantic noise while sharpening the signal on discriminative tokens.
Yelp presents a harder boundary, requiring a higher budget ($r\approx$0.5) to maintain parity (HR@10=0.744 vs. 0.698 on baseline). This is because Yelp reviews are significantly longer (mean 178 tokens vs. $\sim$80 for Amazon) and multi-faceted; aggressive reuse can blur fine-grained sentiment shifts (e.g., ``great food'' vs. ``slow service''), necessitating more online recomputation to preserve ranking order.

\emph{Second}, compared to CacheBlend and EPIC, Table~\ref{tab:accuracy_comparison} shows that \sysname{} consistently dominates generic baselines. 
On Amazon, while RcLLM improves NDCG@10 to 0.438, CacheBlend drops to 0.360 (-10\% vs. baseline) and EPIC collapses to 0.241 (-40\%). Similarly, on Goodreads, RcLLM matches the baseline HR@10 of 0.776, whereas CacheBlend and EPIC drop to 0.653 and 0.551, respectively.
This performance chasm validates our domain-specific design: CacheBlend treats retrieved chunks as unstructured context rather than positional sequence components, disrupting the fine-grained relative scoring needed for ranking 20+ items; while EPIC decouples positional embeddings to allow flexible reuse, but recommendation prompts rely on strict ordering (e.g., history$\rightarrow$item $A$$\rightarrow$item $B$). In contrast to these two baselines, \sysname{} maintains the ranking signal by explicitly preserving the structural skeleton (instruction tokens) and correcting semantic drift where attention is high (heavy hitters).

\subsection{Ablation Studies}\label{sec:ablation}
We perform controlled ablations to isolate the impact of our three core system components: distributed sharding, affinity scheduling, and recomputation budgets.

\subsubsection{Scalability vs. Locality Trade-off}
\label{sec:ablation-instance}

We vary the cluster size $K \in \{1, 20, 40, 80, 100\}$ to evaluate how sharding impacts system performance. Figure~\ref{fig:instance-count} shows the performance speedup obtained by \sysname{} compared to Prefix-Cache, the industrial standard. We make the following observations.

RcLLM consistently outperforms Prefix-Cache across all scales, delivering up to 3.5$\times$ (Qwen3-8B) and 9.5$\times$ (Qwen-72B) speedups in P99 TTFT.
While the absolute speedup narrows slightly as $K$ increases (due to reduced probability of local hits), the system remains robust.
This ablation validates the feasibility of our design: on Amazon, the per-replica cached item footprint drops from $\sim$66M tokens at $K=1$ to $\sim$1.7M tokens at $K=40$ (Figure~\ref{fig:item-storage}), making CPU-side caching practical.
Although sharding reduces the theoretical maximum hit rate (Figure~\ref{fig:hit-rate}), our similarity-aware placement algorithm recovers a substantial fraction of locality by grouping correlated items, maintaining high hit rates even at large $K$.

\subsubsection{Scheduling policy}
\label{sec:ablation-scheduling}

To validate the global scheduler, we compare it against single-objective policies: i) Hit-Only (optimizing purely for cache locality) and ii) Load-Only (optimizing purely for queue depth).
Figure~\ref{fig:scheduling} presents the performance results under different scheduling policies with increasing load (QPS).

As shown in the figure, \sysname{}’s cache-aware routing achieves the lowest (or near-lowest) mean TTFT across all traffic intensities. At high concurrency, Hit-Only degrades sharply (mean TTFT increases by 1.6$\times$--4.0$\times$ at the highest QPS) because it blindly routes requests to popular hot shards, creating severe queueing backlogs.
Load-Only avoids queues but sacrifices locality (hit rate collapses to $<11\%$ at high QPS), forcing frequent on-the-fly recomputation and undermining caching benefits.
In contrast, \sysname{}’s affinity score successfully navigates this Pareto frontier. By dynamically weighting load ($\beta$) against locality ($\alpha$), it exploits cached states during quiet periods while shedding traffic to less-loaded (but colder) nodes during bursts, preventing both compute and queueing bottlenecks.

\subsubsection{Recompute ratio vs.\ latency}
\label{sec:ablation-recompute}

Finally, we quantify the latency penalty of increasing the recomputation budget $r$ in the distributed setting ($K=40$).
Figure~\ref{fig:recompute-latency} presents the CDF of TTFT latency under varying recomputation budgets ($r_{rev}=r_{item}=r$), illustrating the trade-off between fidelity and speed. As $r$ increases from 0.1 to 0.8, the CDF curves shift progressively to the right, indicating a uniform increase in service time across all percentiles.

Despite this penalty, the gap between \sysname{} and the baseline remains decisive. At the empirically optimal budget of $r=0.3$ (orange curve), \sysname{} still substantially outperforms Prefix-Cache (grey dashed). For example, on Amazon with Qwen3-8B (Figure~\ref{fig:amazon-recompute-8b}), P90 TTFT drops from 0.38s (Prefix-Cache) to 0.18s (\sysname{}).

\section{Related Work}

\textbf{LLM-based Recommendation.}
LLM-based recommendation casts ranking and next-item prediction as language generation, covering zero-shot ranking~\cite{wang2023zeroshotnextitemrecommendationusing,DBLP:conf/ecir/HouZLLXMZ24}, prompting and empirical frameworks~\cite{xu2025tapping}, training-free and two-stage ranking~\cite{lee2025starsimpletrainingfreeapproach,yue2023llamarectwostagerecommendationusing}, reasoning-enhanced sequential recommendation~\cite{wang2023drdtdynamicreflectiondivergent}, interactive/explainable recommenders~\cite{gao2023chatrecinteractiveexplainablellmsaugmented}, and direct generation~\cite{ji2023genreclargelanguagemodel}. Recent work further explores agentic recommenders with tool use and multi-step planning~\cite{DBLP:conf/sigir/ZhaoWWTWR24,wang-etal-2024-recmind,zhang2023agentcfcollaborativelearningautonomous,zhang2024generativeagentsrecommendation}, knowledge-augmented ranking with structured sources~\cite{azizi2025llamareclkgragsinglepasslearnableknowledge}, and Transformer architectures tailored to recommendation~\cite{zhou2025onerectechnicalreport,zhai2024actionsspeaklouderwords,chen2024hllmenhancingsequentialrecommendations}. These works improve recommendation quality and modeling capacity; RcLLM is complementary, targeting serving latency for long, non-contiguous recommendation prompts.

\textbf{KV Caching for Accelerating LLM Inference.}
LLM serving systems such as vLLM~\cite{vllm2023} and SGLang~\cite{zheng2024sglang} accelerate inference through exact-prefix caching, but recommendation prompts often append permuted item descriptions after unique user histories, yielding few prefix hits. BAT~\cite{sun2026bat} accelerates recommendation prefill with bilateral attention, yet its reuse pattern remains tied to contiguous structures. CacheBlend~\cite{yao2025cacheblend} and EPIC~\cite{hu2025epic} enable KV sharing across shifted contexts, but they are general-purpose and do not jointly address ranking-sensitive accuracy recovery, locality-aware routing, or massive sharded item catalogs. RcLLM addresses these recommendation-specific serving constraints with stratified caches, affinity scheduling, and selective recomputation.


\section{Conclusion}
\label{sec:conclusion}

This paper presents RcLLM, a distributed serving system that accelerates LLM-based recommendation by reusing non-contiguous KV blocks from semantic histories and candidate items. RcLLM combines stratified caching, similarity-aware item placement, affinity scheduling, and selective recomputation to reduce redundant prefill while preserving ranking fidelity. Across three datasets and two model scales, RcLLM substantially lowers TTFT over prefix caching and keeps accuracy competitive, making real-time generative recommendation practical under industrial latency SLOs.


\section*{Acknowledgements}
This work is supported by Guangdong and Hong Kong Universities ``1+1+1'' Joint Research Collaboration Scheme (project No. 2025A0505000012). Amelie Chi Zhou is the corresponding author.

\bibliographystyle{IEEEtran}
\bibliography{ref}

\end{document}